\documentclass[prd,superscriptaddress,showpacs,floatfix,%
preprintnumbers, nofootinbib,hyperref]{revtex4}  
\usepackage{graphicx}

\def\P{{\mathcal P}}

\begin{document}
\preprint{LAPTH-026/12}
\title{Light sterile neutrino production in the early universe with dynamical neutrino asymmetries}

\author{Alessandro Mirizzi} 
\affiliation{II Institut f\"ur Theoretische Physik, Universit\"at Hamburg, Luruper Chaussee 149, 22761 Hamburg, Germany}

\author{Ninetta Saviano} 
\affiliation{II Institut f\"ur Theoretische Physik, Universit\"at Hamburg, Luruper Chaussee 149, 22761 Hamburg, Germany} 
\affiliation{Dipartimento di Scienze Fisiche, Universit{\`a} di Napoli Federico II,
Complesso Universitario di Monte S. Angelo, I-80126 Napoli, Italy}

\author{Gennaro Miele}
\affiliation{Dipartimento di Scienze Fisiche, Universit{\`a} di Napoli Federico II,
Complesso Universitario di Monte S. Angelo, I-80126 Napoli, Italy}
\affiliation{Istituto Nazionale di Fisica Nucleare - Sezione di Napoli,
Complesso Universitario di Monte S. Angelo, I-80126 Napoli, Italy}

\author{Pasquale Dario Serpico} 
\affiliation{LAPTh, Univ. de Savoie, CNRS, B.P.110, Annecy-le-Vieux F-74941, France}


\begin{abstract}
Light sterile neutrinos mixing with the active ones have been recently proposed
to solve different anomalies observed in short-baseline oscillation experiments. 
These neutrinos can also be  produced by oscillations of the active neutrinos in the early universe, leaving possible
traces on different cosmological observables. Here we perform an updated study
of the neutrino kinetic equations in (3+1) and (2+1) oscillation schemes, dynamically evolving primordial
 asymmetries of active neutrinos and taking into account for the first time CP-violation effects. 
In the absence of neutrino asymmetries, eV-mass scale sterile neutrinos would be completely
thermalized creating a tension with respect to the CMB, LSS and BBN data. 
In the past literature, active neutrino asymmetries have been invoked as a way to  inhibit the sterile neutrino production
via the in-medium suppression of the sterile-active mixing angle. However, neutrino asymmetries also permit
a {\it resonant} sterile neutrino production. We find that if the active species have equal asymmetries $L$, a value $|L|=10^{-3}$ is required
to start suppressing the resonant sterile production, roughly an order of magnitude larger  than what previously expected. When active species have opposite asymmetries the sterile abundance is further enhanced, requiring an even larger $|L|\simeq 10^{-2}$ to start suppressing their production. In the latter case,
CP-violation (naturally expected) further exacerbates the phenomenon. Some consequences for cosmological observables are briefly discussed: for example,
it is likely that moderate suppressions of the sterile species production are associated with significant spectral distortions of
the active neutrino species, with potentially interesting phenomenological consequences especially for BBN. 
\end{abstract}


\maketitle

\section{Introduction}

In  recent years a 
renewed attention has been paid
to light ($m\sim\mathcal O(1)\,$eV) sterile neutrinos mixing with the active ones (see~\cite{Abazajian:2012ys} for a recent review). 
In particular, sterile neutrinos have been proposed to solve different
anomalies observed in short-baseline neutrino experiments,
notably in the ${\overline \nu}_\mu \to {\overline\nu}_e$ oscillations in
LSND~\cite{Aguilar:2001ty} and MiniBoone~\cite{AguilarArevalo:2010wv} experiments, and in ${\overline\nu}_e$ and $\nu_e$ disappearance
revealed by the Reactor Anomaly~\cite{Mention:2011rk} and the Gallium Anomaly~\cite{Acero:2007su}, respectively. 
Scenarios with one (dubbed ``3+1'') or two (``3+2'') sub-eV sterile neutrinos~\cite{Akhmedov:2010vy,Kopp:2011qd,Giunti:2011gz,Giunti:2011hn,Donini:2012tt} have been proposed
to fit the different data. 

Cosmology provides an important arena to test these scenarios. In fact, neutrinos are abundantly produced via weak interactions in the hot (temperature $T\gg 1\,$MeV)  primordial cosmic soup. The mostly sterile mass eigenstate(s) can be produced via oscillations and modify cosmological observables~\cite{Dolgov:1980cq,Barbieri:1989ti,Barbieri:1990vx,Enqvist:1990ad}.  If these additional
states are produced well before (active) neutrino collisional decoupling, they acquire quasi-thermal distributions and behave as extra degrees of freedom 
at the time of big bang  nucleosynthesis (BBN). This would   anticipate  weak interaction decoupling and leading to a larger neutron-to-proton ratio, eventually resulting into a larger $^4$He fraction.  
The non-electromagnetic cosmic radiation content is usually expressed in terms of the effective numbers of thermally excited neutrino species $N_{\rm eff}$. 
The Standard Model (plus active neutrino oscillations) expectation for this parameter is $N_{\rm eff}=3.046$~\cite{Mangano:2005cc}, a result which is only marginally modified even accounting for new interactions between neutrinos and electrons parameterized by four-fermion operators of dimension six allowed by laboratory constraints~\cite{Mangano:2006ar}. If the additional degrees of freedom are still relativistic at the time of cosmic microwave background (CMB) formation, 
the same parameter $N_{\rm eff}$ can be constrained by a detailed study of the CMB angular power spectrum, especially when combined with other cosmological probes. Quite an excitement (see e.g.~\cite{Hamann:2010bk,GonzalezGarcia:2010un}) has been stimulated by the result in the current best fit of WMAP, SDSS II-Baryon Acoustic Oscillations and Hubble Space Telescope data, yielding a $68\%$ CL range on $N_{\rm eff}=4.34^{+0.86}_{-0.88}$~\cite{Komatsu:2010fb} in the assumption of a $\Lambda$CDM universe. Once accounting for the parameter degeneracy in the determination of the first angular peak properties, which can be adjusted by choosing different combinations of other cosmological parameters,  it turns out that such results are almost completely due to the large-$\ell$, damping tail of the CMB spectrum, as even more clear when adding ACBAR~\cite{Reichardt:2008ay} and ACT~\cite{Das:2010ga} small scale data (see~\cite{Hou:2011ec} for a pedagogical account).
The reliability of BBN constraints is plagued by systematics in the determination of primordial abundances. Yet, within conservative but reasonable assumptions, standard BBN
calculations do not allow for  $N_{\rm eff}$ larger than about 4.1 at 95\% CL~\cite{Mangano:2011ar}
with only  a weak, statistically non-significant preference for a larger-than-standard value of   $N_{\rm eff}$. In order to allow for sufficiently large effects at CMB epoch while accommodating BBN constraints, some authors have envisaged for example the introduction of relatively large neutrino-antineutrino asymmetries. This in order to (partially) compensate for the effect of $N_{\rm eff}$  on $^4$He by a counter-effect due to $\nu_e-\overline\nu_e$ distributions in weak rates~\cite{Hamann:2011ge}. On the other hand, the CMB preference 
for a large  $N_{\rm eff}$ usually comes with a price: for example, a tension with cluster determination of dark matter abundance has been noticed in~\cite{Hou:2011ec}. Also, the basic tenet that these mostly sterile $\nu$'s behave essentially as radiation at CMB epoch is untenable: laboratory data require one or more relatively massive ($m\sim 1\,$eV) extra states. CMB data in combination with LSS ones are a particularly sensitive probe of massive neutrinos (for a review see e.g.~\cite{Lesgourgues:2006nd,Wong:2011ip}). At face value, when accounting for the fact that these extra species are massive, the scenario hinted at by laboratory data is actually {\it disfavored} by cosmology, unless rather radical and contrived cosmological model modifications are introduced~\cite{Dodelson:2005tp,Joudaki:2012uk, Hamann:2011ge,GonzalezGarcia:2010un,Giusarma:2011zq}.

Given this partially contradictory situation and the existing laboratory anomalies, it is of paramount importance to study the physical conditions under which the sterile neutrino production actually takes place, as preliminary step to any phenomenological consideration.  As already mentioned, sterile neutrinos are produced in the early universe by the mixing with the active species. Therefore,
 in order to determine their abundance  it is necessary to solve the quantum kinetic equations describing
 the active-sterile oscillations~\cite{Sigl:1992fn,McKellar:1992ja}. This problem has been studied in a long series of papers
(see e.g.~\cite{Enqvist:1990ek,Enqvist:1991qj,Enqvist:1990dq,DiBari:1999ha,DiBari:1999vg,DiBari:2000wd,%
DiBari:2001ua,Dolgov:1999wv,DiBari:2000tj,Foot:1995bm,Foot:1995qk,Bell:1998ds,Kirilova:1997sv,Kirilova:1999xj,%
Abazajian:2004aj,Kishimoto:2006zk,Dolgov:2003sg,Cirelli:2004cz,Chu:2006ua,Abazajian:2008dz,Melchiorri:2008gq,Hannestad:2012ky}), finding 
 a broad range of possible outputs depending on the sterile neutrino masses and mixing angles with the active species. 
Since the solution of the non-linear neutrino evolution equations  is numerically challenging,  different approximations
have been adopted. In particular, most of the previous studies solve the equations
in a simplified (1+1) scenario in which only a (mostly) active neutrino mixes with a (mostly) sterile one. 
Recently,  also multi-flavor cases have been 
presented~\cite{Dolgov:2003sg,Melchiorri:2008gq}. In particular, active-sterile oscillations in a (3+2) scheme
have been studied~\cite{Melchiorri:2008gq}. It has been found that, for the mass and mixing parameters needed to describe the short-baseline anomalies,
in the standard scenario sterile neutrinos would be completely thermalized in the early universe, creating an unwelcome tension
with cosmological observations as mentioned above.

On the other hand, a possible escape-route to reconcile sterile neutrinos  with cosmological data consists in  the inclusion of
a  primordial  asymmetry between neutrinos and antineutrinos~\cite{Foot:1995bm}
\begin{equation}
L = \frac{n_{\nu}-n_{\bar\nu}}{n_{\gamma}} \,\ .
\end{equation}
In principle, one would expect the lepton asymmetry to be of the same order of magnitude of the baryonic one, 
$\eta = (n_B-n_{\bar B})/n_{\gamma} \simeq 6 \times 10^{-10}$. 
Indeed, to respect the charge neutrality the asymmetry in the charged leptons must match the above number to
a high degree.
However, since neutrinos are neutral  the constrains on $L$ are quite loose,
allowing also  $|L|\simeq 10^{-2}-10^{-1}$~\cite{Dolgov:2002ab,Serpico:2005bc,Pastor:2008ti,Mangano:2010ei,DiValentino:2011sv,Mangano:2011ip,Castorina:2012md}. 
Moreover, there are models for producing large $L$ and small
 $\eta$~\cite{Harvey:1981cu,Casas:1997gx,Dolgov:2002wy}.

A  neutrino asymmetry implies an additional ``matter term potential'' in the equations of motion. If sufficiently
large, one expects this term to block the active-sterile flavor conversions via the in-medium suppression of the mixing angle.
However, this term can also generate  Mikheev-Smirnov-Wolfenstein~\cite{Matt} (MSW)-like resonant flavor conversions  among active and sterile
neutrinos. In particular, in a recent (1+1) study~\cite{Hannestad:2012ky} it has been found that for sterile neutrinos with parameters
preferred by the laboratory hints,  a neutrino asymmetry $L = 10^{-2}$ would strongly suppress their production.
This would  reconcile them with the cosmological observations.    
 In~\cite{Chu:2006ua} the question was addressed  of how large should be the value of $L$  in order to have a significant
reduction of the sterile neutrino abundance. 
The authors solved the equations of motion in   a simplified (3+1) scheme inspired by LSND, finding that  
$L \sim 10^{-4}$ was enough to relieve the tension between sterile neutrinos and cosmology. 
However, this result has to be taken \emph{cum grano salis}. Indeed, the authors fixed the lepton asymmetry as
an initial condition taken constant during the flavor evolution. Nevertheless, this quantity is expected to dynamically
evolve due to the flavor conversions. Moreover, they solved the coupled equations of motion
by effectively reducing the degrees of freedom via the constraint of neglecting resonant transitions between 
sterile and active neutrino species in the antineutrino sector, that would be possible for the negative neutrino
asymmetries they considered. 

 only for neutrinos, neglecting
the antineutrino sector, in which   resonant conversions  with active neutrinos would occur for the negative lepton
asymmetries  they considered.

The purpose of our work is to revisit the thermalization of sterile neutrinos in the early universe in the presence of
primordial neutrino asymmetries, taking a complementary approach to most of other studies. In particular, many investigations
have focused on large scans of the sterile neutrinos parameter space, often neglecting some physical effects in order
to keep the problem computationally manageable. Here, we rather stick to a benchmark set of best fit value parameters
for the sterile state ``suggested'' by laboratory experiments, but we go beyond most approximations used in the previous studies.
In particular, we shall consider (3+1) and (2+1) schemes inspired by the recent fits of  all the short-baseline,
reactor, and solar neutrino anomalies. The inclusion of more than an active neutrino that mixes with the sterile state
allows to explore effects that are not possible in a simplified (1+1) scenario. 
For example, having more than one active neutrino one can study both cases in which the neutrino asymmetries are 
equal or opposite between the active species. Moreover, one can take into account more than one mixing angle between
 the  active and the sterile neutrinos. 
CP violating effects in oscillations thus become a natural possibility, which we also consider here, to the best of our knowledge for the first time in this context~\footnote{In the absence of sterile neutrinos, the impact of CP violation onto cosmological active neutrino asymmetries has been studied in~\cite{Gava:2010kz}.}. 
We decide to devote our work to a detailed study of all these effects still unexplored. We  find that each of them could have a relevant impact in the determination of the final abundance of sterile neutrinos.  Generically, we realize that ``masking'' cosmological consequences of additional sterile neutrinos is harder than previously deduced in simplified treatments.  Here we do not aim at deriving detailed cosmological predictions, rather the correct (and reliable) qualitative physical trend. In this spirit, we can content ourselves with a momentum-averaged description, but  otherwise we deal with the complete problem, i.e. we solve essentially the exact
equations of motion as opposed to simplified ones studied in the past literature.
  
The plan of our paper is as follows. 
In Sec.~II we introduce the (3+1) neutrino mixing framework we will use as benchmark for our study. 
In Sec.~III we describe the active-sterile neutrino flavor evolution in the early universe. In particular,
we introduce the equations of motion for the neutrino ensemble. Then, we present the average-momentum approximation
we use to solve numerically the neutrino evolution. Finally, we compare the different interaction strengths that enter
the equations of motion. 
 In Sec.~IV we present the results of the sterile neutrino production in the (3+1) scenario for different values of
 the primordial neutrino asymmetries, taken equal among the different active flavors. 
In Sec.~V we calculate the sterile neutrino abundance in different cases in a (2+1) scenario, where the active sector
$(\nu_e,\nu_{\mu})$
is associated with the atmospheric mass-square splitting $\Delta m^2_{\rm atm}$ and with the 1-3 mixing angle $\theta_{13}$. 
At first, we take into account both the active-sterile mixing angles $\theta_{es}$ and $\theta_{\mu s}$. In this situation we describe
 both cases with equal and opposite initial neutrino asymmetries among the active species. We also consider the effects
of CP violation in the sterile sector. Then, we present  a case in which only electron neutrinos mix with the sterile ones.
 In Sec.~VI we discuss the impact of the sterile neutrino production on the neutron/proton  ratio ($n/p$) in the early universe. 
Finally, in Sec.~VII we comment about future developments of our study and we conclude.

\section{(3+1) neutrino framework}
 
 Short-baseline neutrino oscillation data have been fitted adding to the usual three active neutrino  species one (3+1 mixing scheme) or two (3+2 mixing scheme)
massive sterile states~\cite{Akhmedov:2010vy,Kopp:2011qd,Giunti:2011gz,Giunti:2011hn}. In spite of the presence of a tension in the interpretation of the data~\cite{Giunti:2011hn}, the 3+1 neutrino mixing scenario is attractive for its simplicity. It is also less likely to lead to a strong exclusion by cosmological arguments. Therefore, in our work  we will consider only this  extension of the three-neutrino mixing framework. In such a four-neutrino mixing scheme, the flavor neutrino basis is composed by the three active neutrinos $\nu_e, \nu_{\mu}, \nu_{\tau}$ and by a sterile neutrino $\nu_s$.  The flavor eigenstates $\nu_\alpha$ are related to the mass eigenstates $\nu_i$ ($i=1,\ldots 4$, ordered by growing mass) via a unitary matrix ${\mathcal U}$ through~\cite{GiuntiKim,Abazajian:2012ys}
\begin{equation}
\nu_{\alpha} = {\mathcal U}_{\alpha i}^* \,  \nu_i \,\ , \,\ \,\ {\mathcal U}{\mathcal U}^{\dagger}=
{\mathcal U}^{\dagger}{\mathcal U}= I \,\ , 
\end{equation}
By neglecting for the moment the presence of arbitrary phases, also responsible for CP violation effects, the matrix $\, {\mathcal U}$  can be parameterized as a product of 4$\times 4$ Euler rotation matrices $R_{ij}$ acting in the  $(i,j)$ mass eigenstate subspace, each characterized by a mixing angle $\theta_{ij}$. Following e.g.~\cite{Maltoni:2001bc}  one can write
\begin{equation}
{\mathcal U}= R_{34}R_{24}R_{23}R_{14}R_{13}R_{12} \,\ ,
\end{equation}
where we order the flavor eigenstates in such a way that if all angles are vanishing we have the correspondence	$(\nu_e,\nu_\mu,\nu_\tau ,\nu_s)=(\nu_1,\nu_2,\nu_3, \nu_4)$. In the limit where the three mixing angles $\theta_{i4}$ vanish, the above matrix reduces to
\begin{equation}
\lim_{\theta_{i4}\to 0}{\mathcal U} = 
\left(\begin{array}{cc} 
U(\theta_{12}, \theta_{13}, \theta_{23}) &0 \\
0 & 1 \\
\end{array}\right)~,
\label{eq:matrix}
\end{equation}
where $U$ is the $3 \times 3$ unitary mixing matrix among  the active neutrinos defined in terms of three rotation angles $\theta_{ij}$, ordered as for the quark
mixing matrix~\cite{Nakamura:2010zzi}.
In the following we fix the values of these three mixing angles to the current best-fit from global analysis of the different active neutrino oscillation data~\cite{Fogli:2012ua} (see also~\cite{Tortola:2012te}), i.e.
\begin{eqnarray}
\sin^2 \theta_{12} &=& 0.307 \,\ , \nonumber \\
 \sin^2 \theta_{23} &=& 0.398 \,\ , \nonumber \\
 \sin^2  \theta_{13} &=& 0.0245 \,\ .
\end{eqnarray}
We remind that the early hints for a  ``large'' value of $\theta_{13}$, suggested by the long-baseline $\nu_{\mu}$-$\nu_e$ experiments~\cite{Abe:2011sj,Adamson:2011qu} and Double Chooz reactor experiment~\cite{Abe:2011fz} in combination with the global analysis of  the neutrino data~\cite{Fogli:2008jx,Fogli:2011qn}, have been recently confirmed by the measurement of 
the Daya Bay~\cite{An:2012eh} and Reno~\cite{Ahn:2012nd}  reactor experiments. We neglect CP violating effects in the active sector. The corrections to the $3\times 3$ ``active'' neutrino sub-matrix of ${\mathcal U}$ is only second order in the mixing angles of the sterile state. We shall assume (as also done in phenomenological studies)
that at most two mixings of the fourth neutrino to the three active ones are non vanishing, namely we put  ${\mathcal U}_{\tau 4}=0$. We take from fits of the short-baseline data the values~\cite{Giunti:2011gz}
\begin{eqnarray}
\sin\theta_{es} \simeq |{\mathcal U}_{e4}| =\sqrt{0.025} \,\ , \nonumber \\
\sin\theta_{\mu s} \simeq |{\mathcal U}_{\mu 4}| =\sqrt{0.023} \,\ ,
\label{eq:stermix}
\end{eqnarray}
where the equality holds (in our parameterization) up to corrections quadratic in the mixing angles of the fourth state. It is worth noting that the ``sterile'' mixing angles
result of the same order of $\theta_{13}$. Any quantitative interpretation of ``anomalies'' in terms of mixing with steriles should thus be made to account for oscillations among the
active states.

The 4$\nu$ mass spectrum is parameterized as~\cite{Fogli:2005cq}
\begin{equation}
{\mathcal M}^2 = \textrm{diag}(m_1^2, m_2^2, m_3^2, m_4^2) = \textrm{diag}\left(-\frac{1}{2}\Delta m^2_{\rm sol} \,\ , +
\frac{1}{2} \Delta m^2_{\rm sol} \,\ , 
   \Delta m^2_{\rm atm}  \,\ , \Delta m^2_{\rm st}  \right) \,\ ,  
\end{equation}
where the solar and the atmospheric mass-square differences are given  by~\cite{Fogli:2012ua}
\begin{eqnarray}
\Delta m^2_{\rm sol}/\textrm{eV}^2 = 7.54 \times 10^{-5} \,\ , \nonumber \\
\Delta m^2_{\rm atm}/\textrm{eV}^2 = 2.43 \times 10^{-3}  \,\ ,
\end{eqnarray}
respectively. Note that here and throughout we  assume  normal mass hierarchy,  i.e. $\Delta m^2_{\rm atm}>0$. The sterile-active mass splitting from the short-baseline fit in the 3+1 model is given 
by~\cite{Giunti:2011gz}
\begin{equation}
\Delta m^2_{\rm st} /\textrm{eV}^2 = 0.89 \,\ .
\end{equation}
Therefore, it results in a clear hierarchy among the mass differences, i.e.  $\Delta m^2_{\rm st} \gg \Delta m^2_{\rm atm} \gg \Delta m^2_{\rm sol}$.  When restricting to the (2+1) cases, no $4\times 4$ formalism is needed. The conventions are formally equivalent to the familiar three active neutrino mixing angles, as well as the standard parameterization of the Dirac CP phase~\cite{Nakamura:2010zzi}, with only the numerical values for the mixing and  mass splittings to be changed (see Sec.~\ref{2plus1} for details). 
\section{Neutrino flavor evolution   in the early universe}

\subsection{Equations of Motion}

Following~\cite{Dolgov:2002ab},  in order to describe the time evolution of $\nu-\overline\nu$ ensemble  in the early universe, it proves useful to define the following dimensionless variables which replace time, momentum and photon temperature, respectively
\begin{equation}
x \equiv m\,a  \qquad  y \equiv p\,a  \qquad z \equiv T_\gamma\,
a~, \label{comoving}
\end{equation}
where $m$ is an arbitrary mass scale which can be put e.g. equal to 1 MeV. Note that the function $a$ is normalized, without loss of generality, so that $a(t)\to 1/T$ at large
temperatures, $T$ being the common temperature of the particles in equilibrium far from any entropy-release process. With this choice, $a^{-1}$ can be identified with the initial temperature of thermal, active neutrinos. 

In order to  characterize the active-sterile neutrino oscillations we describe the neutrino (antineutrino) ensemble in terms of  $4\times 4$ density matrices $\varrho$ ($\bar\varrho$)\footnote{Note that in natural units $\varrho$  and $\bar\varrho$ are dimensionless variables.}
\begin{equation}
\varrho(x,y) =
\left(\begin{array}{cccc} 
\varrho_{ee} &  \varrho_{e\mu} & \varrho_{e\tau} & \varrho_{es} \\
\varrho_{\mu e}  & \varrho_{\mu \mu} & \varrho_{\mu \tau} & \varrho_{\mu s} \\
\varrho_{\tau e} & \varrho_{\tau \mu} &  \varrho_{\tau \tau} & \varrho_{\tau s} \\
\varrho_{s e} &\varrho_{s \mu} &\varrho_{s \tau}& \varrho_{ss}
\end{array}\right)~. 
\label{eq:rho}
\end{equation}
In terms of $\varrho$  and $\bar\varrho$ the Equations of Motion (EoMs) for the neutrino ensemble assume the form~\cite{McKellar:1992ja,Sigl:1992fn,Dolgov:2002ab}
\begin{eqnarray}
i  \frac{d\varrho}{dx} &=& + \frac{x^2 }{2 m^2\,y\,\overline{H}} \left[{\mathcal U}^{\dagger}{\mathcal M}^2{\mathcal U}, \varrho \right]
+  \frac{\sqrt{2} G_F \, m^2}{x^2 \, \overline{H}}\left[\left(-\frac{8\, y\, m^2}{3\, x^2 \, m_W^2} {\sf E_\ell}-\frac{8\, y\, m^2 }{3\, x^2 \,  m_Z^2} {\sf E_\nu} + {\sf N}_\nu  \right),\varrho \right] \nonumber \\
&+& \frac{x \, C[\varrho] }{m \, \overline{H}} \label{eq:eomrho} \,\ , \\
i \frac{d\bar\varrho}{dx}  &=& - \frac{x^2}{2 m^2\,y\, \overline{H}} \left[{\mathcal U}^{\dagger}{\mathcal M}^2{\mathcal U}, {\bar\varrho} \right]
+ \frac{\sqrt{2} G_F \, m^2}{x^2 \, \overline{H}} \left[\left(+\frac{8\, y\, m^2}{3\, x^2 \,  m_W^2} {\sf E_\ell}+\frac{8\, y\, m^2 }{3 \, x^2 \, m_Z^2} {\sf E_\nu} + {\sf N}_\nu \right), {\bar\varrho} \right] \nonumber \\
&+& \frac{x \, C[{\bar\varrho}]}{m \, \overline{H}}  \,\ ,  \label{eq:eombarrho}\\
x\frac{d \varepsilon}{d x} &=&\varepsilon-3\P \,\ . \,\,\,\,\,\,\,\,\,\,\,\,\,\,\,\,\,\,\,\,\,\,\,\,\,\,\,\,\,\,\,\,\,\,\,\,\,\,\,\,\,\,\,\,\,\,\,\,\,\,\,\,\,\,\,\,\,\,\,\,\,\,\,\,\,\,\,\,\,\,\,\,\,\,\,\,\,\,\,\,\,\,\,\,\,\,\,\,\,\,\,\,\,\,\,\,\,\,\,\,\,\,\,\,\,\,\,\,\,\,\,\,\,\,\,\,\,\,\,\,\,\,\,\,\,\,\,\,\,\,\,\,\,\,\,\,\,\,\,\,\,\,\,\,\,\,\,\,\,\,\,\,\,\,\,\,\,\,\,\,\,\,\,\,\,\,\,\,\,\,\,\,\,\,\,\,\
\label{eq:eomz}
\end{eqnarray}
In the previous expressions $\overline{H}$ denotes the properly normalized Hubble parameter, namely
\begin{equation}
\overline{H} \equiv \frac{ x^2}{m} H = \frac{ x^2}{m} \sqrt{\frac{8\pi \, \epsilon(x,z(x))}{3 \, M_{Pl}^2}}=\left(\frac{m}{M_{Pl}}\right) \sqrt{\frac{8\pi \varepsilon(x,z(x))}{3}} \,\ ,
\end{equation} 
where the total energy density and pressure of the plasma, $\epsilon$ and $P$, enter through their ``comoving transformed'' $ \varepsilon \equiv  \epsilon(x/m)^4$ and $\P \equiv P(x/m)^4$ respectively. Since for most of the temperatures we are interested in, electrons and positrons are the only charged leptons populating the plasma in large numbers, to a very good approximation the total energy density can be expressed as the sum 
\begin{equation}
\varepsilon(x,z(x))\simeq \varepsilon_{\gamma}+\varepsilon_{e}+\varepsilon_{\nu}\,,
\end{equation} 
where
\begin{eqnarray}
\varepsilon_\gamma&=&\frac{\pi^2}{15}z^4(x)\,,\\
\varepsilon_e&=&\frac{1}{\pi^2}\int_0^\infty dy\,y^3 \,  \left[ f_{FD}(y/z(x)-\phi_e)+f_{FD}(y/z(x)+\phi_e)\right]
\simeq \frac{7 \, \pi^2}{60} \, z^4(x)
\,,\\
\varepsilon_{\nu}&=& 
 \frac{1}{2 \pi^2} \int   \, dy \, y^3
 \textrm{Tr}[\varrho(x,y) +\bar\varrho(x,y)] 
\equiv\frac{7}{8}\frac{\pi^2}{15}N_{\rm eff}\,.\label{neff}
\end{eqnarray} 
Note that due to the range of temperature $T$ considered we have safely assumed massless $e^\pm$ that, due to the fast electromagnetic interactions, have a Fermi-Dirac distribution $f_{FD}(y/z(x) \mp \phi_e) \equiv 1/(\exp(y/z(x) \mp \phi_e)+1)$ respectively. The reduced electron chemical potential  $\phi_e$ is in principle a dynamical variable that requires a further equation (the electric charge conservation) in order to be evolved consistently. However, for our purpose electrons are only important when their energy density is dominated by pairs, rather than by the $e^-$ excess due to the baryon asymmetry, and $\phi_e$ can be put equal to zero.

The first term on the r.h.s. of the EoMs (\ref{eq:eomrho}) and (\ref{eq:eombarrho}) is responsible for the vacuum neutrino oscillations.  In the second term,  the diagonal matrix ${\sf E_\ell}$  related to the energy density of charged leptons under the previous assumptions takes the form
\begin{equation}
{\sf E_\ell} \equiv \textrm{diag}(\varepsilon_{e}, 0,0,0) = \textrm{diag}
\left(\frac{7 \, \pi^2}{60} \, z^4(x), 0,0,0\right) \,\ .
\label{eq:Ell}
\end{equation}
Moreover we have 
 \begin{eqnarray}
 {\sf N}_\nu &=&  \frac{1}{2 \pi^2} \int  dy \,  y^2 \,
 \{ {\sf G}_s (\varrho(x,y) -{\bar\varrho}(x,y)){\sf G}_s + {\sf G}_s \textrm{Tr} \left[(\varrho(x,y) -{\bar\varrho}(x,y)){\sf G}_s\right]  \}
 \,,\label{eq:neutrinodens}  \\
  {\sf E}_\nu& = &  \frac{1}{2 \pi^2} \int   dy \, y^3 \,
 {\sf G}_s (\varrho(x,y) +\bar\varrho(x,y)){\sf G}_s  \,.
 \label{eq:neutrinodens2}
 \end{eqnarray}
These terms make  the EoMs non-linear and are the main numerical challenge in dealing with this physical system. Note that the matrix ${\sf N}_\nu$ is related to the {\it difference} of the  density matrices of neutrinos and antineutrinos, while ${\sf E}_\nu$ is related to their sum. The matrix ${\sf G}_s =\textrm{diag}(1,1,1,0)$  in flavor space contains the dimensionless coupling constants. We remark that in the presence of more than one active species,  the ${\sf N}_\nu$ matrix also contains off-diagonal terms.   
The last term at r.h.s. of Eqs.~(\ref{eq:eomrho}) and (\ref{eq:eombarrho}) is the collisional term proportional to $G_F^2$.  We will present an approximate expression of this term in the Sec.~\ref{SMapprox}.  Finally, Eq. (\ref{eq:eomz}) basically provides an evolution equation for the quantity $z(x)$. On the other hand, even when a fourth neutrino
is populated in the plasma via early oscillations (before the active neutrino decoupling), the correction with respect to the initial value of $\varepsilon$
is at most of the order of $\sim 10\%$. Consistently with other papers in this field we neglect this small effect in this exploratory study, although we note that it should be accounted for in more accurate predictions of cosmological observables. This also implies that we can keep $z$ as constant and equal to 1, which is a good approximation in the epoch considered in this work. As a consequence, $\varepsilon$ is not dynamical and assumes a numerical value almost equal to 3.54.
A final comment is that we omit the familiar refractive term due to the matter asymmetry (i.e. $\propto (n_{e^-}-n_{e^+})\propto \eta$) from the EoMs. Albeit, it may, in principle induce resonances, for the parameters of interest here the resonance would fall at such high temperature that the system is still collisionally dominated and no coherent process is effectively taking place.

\subsection{``Average momentum'' approximation}\label{SMapprox}
In the presence of continuous neutrino momentum distributions, to solve the full  set of EoMs (\ref{eq:eomrho}) and (\ref{eq:eombarrho}) turns out to be a computationally demanding task. In order to perform a more treatable numerical study of the flavor evolution  for different neutrino asymmetries, which is able to catch the main features of the more involved complete computation,  in the following we will restrict ourselves to a average-momentum approximation, based on the Ansatz (and similarly for antineutrinos)
\begin{equation}
\varrho(x,y)\to f_{FD}(y)\,\rho(x)\,.
\label{eq:approxeq}
\end{equation}
By this assumption, and in absence of asymmetries, at equilibrium one would simply  get $\rho={\sf I}$. In terms of  
Eq.~(\ref{eq:approxeq}) the set of EoMs (\ref{eq:eomrho}) and (\ref{eq:eombarrho}) can be rewritten as
\begin{eqnarray}
i  \frac{d\rho}{dx} &=& + \frac{x^2 }{2 m^2 \,\overline{H}} \left\langle \frac{1}{y} \right\rangle \left[{\mathcal U}^{\dagger}{\mathcal M}^2{\mathcal U}, \rho \right]
+  \frac{\sqrt{2} G_F \, m^2}{x^2 \, \overline{H}}\left[\left(-\frac{8 \langle y \rangle m^2}{3\, x^2 \, m_W^2} {\sf E_\ell}-\frac{8 \langle y \rangle \, m^2 }{3\, x^2 \,  m_Z^2} {\sf E_\nu} + {\sf N}_\nu  \right),\rho \right] \nonumber\\
&+& \frac{\widehat{C}[\rho]  }{x^4 \, \overline{H}} \label{eq:eomrho_bis} \\
i \frac{d\bar\rho}{dx}  &=& - \frac{x^2}{2 m^2 \,\overline{H}}\left\langle \frac{1}{y} \right\rangle \left[{\mathcal U}^{\dagger}{\mathcal M}^2{\mathcal U}, {\bar\rho} \right]
+ \frac{\sqrt{2} G_F \, m^2}{x^2 \, \overline{H}} \left[\left(+\frac{8 \langle y \rangle m^2}{3\, x^2 \,  m_W^2} {\sf E_\ell}+\frac{8 \langle y \rangle m^2 }{3 \, x^2 \, m_Z^2} {\sf E_\nu} + {\sf N}_\nu \right), {\bar\rho} \right] \nonumber\\
&+& \frac{\widehat{C}[{\bar\rho}]}{x^4 \, \overline{H}}\,, \label{eq:eombarrho_bis}
\end{eqnarray}
where by definition $\langle g(y)\rangle \equiv \int_0^\infty y^2 \, g(y) \,  f_{FD}(y)  \, dy / \int_0^\infty y^2  \, f_{FD}(y) \, dy$. According to this notation $\langle y \rangle = 3.15$ and $\langle 1/y \rangle = 0.456 \neq 1/\langle y \rangle$. The non-linear terms ${\sf N}_\nu$ and  ${\sf E}_\nu$ of Eqs.~(\ref{eq:neutrinodens},\ref{eq:neutrinodens2})
assume the form
\begin{eqnarray}
{\sf N}_\nu &=& \frac{3\,\zeta(3)}{4\pi^2} 
 \{ {\sf G}_s (\rho -{\bar\rho}){\sf G}_s + {\sf G}_s \textrm{Tr} \left[(\rho -{\bar\rho}){\sf G}_s\right]  \}\,, \\
{\sf E}_\nu &=&  \frac{7}{8}\frac{\pi^2}{30} {\sf G}_s (\rho +\bar\rho){\sf G}_s\,.
\end{eqnarray}
Note that  only the ``active'' 3$\times 3$ submatrix of the whole density matrix enters the above interaction terms. Moreover,  by using the approximate form~\cite{Chu:2006ua} for the collisional terms in Eqs~(\ref{eq:eomrho}) and (\ref{eq:eombarrho}), we get the expressions 
\begin{eqnarray}
 \widehat{C}[{\rho}] &=& -\frac{i}{2} G_F^2 \, m^4 
 (\{{\sf S}^2, \rho-{\sf I}\} 
 - 2 {\sf S}(\rho-{\sf I}){\sf S} + \{{\sf A}^2,
 (\rho-{\sf I})\} + 2 {\sf A}(\bar\rho-{\sf I}){\sf A}) \,\ , \nonumber\\
  \widehat{C}[{\bar\rho}] &=& -\frac{i}{2} G_F^2 \, m^4 
 (\{{\sf S}^2, \bar\rho-{\sf I}\} 
 - 2 {\sf S}(\bar\rho-{\sf I}){\sf S} + \{{\sf A}^2,
 (\bar\rho-{\sf I})\} + 2 {\sf A}(\rho-{\sf I}){\sf A}) \,\ .
\end{eqnarray}
The above expressions have been derived assuming null neutrino lepton asymmetries. However, since in our study we will restrict ourselves to $|L| \leq 10^{-2}$,
the correction induced would be negligible (see, e.g.,~\cite{Bell:1998ds}).   In flavor space, the matrices  ${\sf S, A}$ write ${\sf S} = \textrm{diag}(g_s^e, g_s^\mu, g_s^\tau,0)$ and 
 ${\sf A} = \textrm{diag}(g_a^e, g_a^\mu, g_a^\tau,0)$, respectively, and contain  the numerical coefficients for the scattering and annihilation processes of the different flavors. 
Numerically one finds~\cite{Enqvist:1991qj}
\begin{eqnarray}
(g_s^e)^2 &=& 3.06 \,\ \,\ , \,\ (g_a^e)^2 = 0.50 \,\ , \nonumber \\
(g_s^{\mu,\tau})^2 &=& 2.22 \,\ \,\ , \,\ (g_a^{\mu,\tau})^2 = 0.28 \,\ .
\end{eqnarray}

The initial conditions for the  density matrix $\rho$ are written
\begin{eqnarray}
\rho_{\rm in} &=& \textrm{diag}\left(1+\frac{4}{3} L_{e}, 1+ \frac{4}{3} L_{\mu}, 1+ \frac{4}{3} L_{\tau}, 0\right) \,\ , \nonumber \\
{\bar\rho} _{\rm in} &=& \textrm{diag}\left(1-\frac{4}{3} L_{e}, 1- \frac{4}{3} L_{\mu}, 1- \frac{4}{3} L_{\tau},0\right)\label{43} \,\ ,
\end{eqnarray}
where the neutrino asymmetries in the different flavors 
 are related to dimensionless chemical potentials $\xi_{\nu}= \mu_{\nu}\,a$ through
\begin{equation}
L = \frac{\pi^2}{12 \zeta(3)} \left( \xi_{\nu}
+ \frac{{\xi_{\nu}}^3}{{\pi}^2} 
\right) .
\end{equation}
Note that the expression in Eq.~(\ref{43}) is only valid at leading order in $L$.  

It is convenient to have an estimate of the different dimensionless factors multiplying $\rho$ and $\bar\rho$ on the r.h.s. of 
Eqs.~(\ref{eq:eomrho_bis}) and  (\ref{eq:eombarrho_bis}), respectively.  The vacuum oscillation term is proportional, apart from a matrix  whose coefficients are ${\mathcal O}$(1), to the  quantity
\begin{equation}
\Omega_{\rm vac} \equiv  \frac{x^2 \, \Delta m^2}{2\,\overline{H}\, m^2} \left\langle \frac{1}{y} \right\rangle = 2.3 \times 10^{-13} \,
\left( \frac{\Delta m^2}{{\rm eV}^2}\right) \, \frac{x^2 }{\overline{H}} \, ,
\label{eq:Omvac}
\end{equation}
Taking into account the  $e^+e^-$ pairs only, the matter potential in Eqs.~(\ref{eq:eomrho_bis}) and  (\ref{eq:eombarrho_bis}) except for the different sign for neutrinos and antineutrinos, is proportional to 
\begin{equation}
\Omega_{\rm matt} = \frac{8\sqrt{2}\langle y\rangle G_F \, m^4}{3\,m_W^2}\frac{7\,\pi^2}{60}\left(\frac{1}{x^4 \, \overline{H}}\right)= 2.4 \times 10^{-20}\, 
\frac{1}{x^4 \, \overline{H}}\,\ .
\end{equation}

The neutrino-neutrino interaction strength   gives two terms, respectively proportional to 
\begin{eqnarray}
\Omega_{\rm asy} &=&  \frac{\sqrt{2} G_F\,m^2}{x^2\, \overline{H}} \, \frac{3\,\zeta(3)}{4 \, \pi^2} =  1.5 \times 10^{-12} \, \frac{1}{x^2\, \overline{H}} \,\ ,\nonumber \\
\Omega_{\rm sym} &=& \frac{8\sqrt{2}\langle y\rangle G_F \, m^4}{3\,m_Z^2}\frac{7\,\pi^2}{240}\left(\frac{1}{x^4 \, \overline{H}}\right) =\frac{1}{4}\left(\frac{m_W}{m_Z}\right)^2\Omega_{\rm matt} =\frac{\cos^2\theta_W}{4} \,  \Omega_{\rm matt}\,\ .
\end{eqnarray}
Finally,  the collisional term is proportional to
\begin{equation}
\Omega_{\rm coll} = \frac{ G_F^2 \, m^4}{2 \, x^4 \, \overline{H}}  =   6.8 \times 10^{-23} \,  \frac{1}{x^4 \, \overline{H}} \, .
\end{equation}

In order to get an idea of the strength of the different interaction terms, 
in Figure~1 we plot as a  function of the temperature $T$, $\Omega_{\rm vac}$ (solid curve), $\Omega_{\rm matt}$
(long-dotted curve),  $\Omega_{\rm asy} \times \Delta_e$ (dashed curve), 
$\Omega_{\rm sym} \times \Sigma_{ee}$ (short-dotted curve), 
$\Omega_{\rm coll} \times [(g_s^e)^2 + (g_s^\mu)^2]$  (dash-dotted curve). Here we use as mass square difference $\Delta m^2_{\rm st}$, 
$\Delta_e = 2 (\rho_{ee}-\bar{\rho}_{ee}) + (\rho_{\mu \mu}-\bar{\rho}_{\mu \mu}) +
 (\rho_{\tau \tau}-\bar{\rho}_{\tau \tau}) = 8/3
(2 L_e + L_\mu + L_\tau$), where for illustration we fixed $L_e=L_\mu=L_\tau = 10^{-4}$.
Finally    $ \Sigma_{ee} =  (\rho_{ee}+\bar{\rho}_{ee}) \simeq 2$.  

\begin{figure}[!t]  
\includegraphics[angle=0,width=0.8\columnwidth]{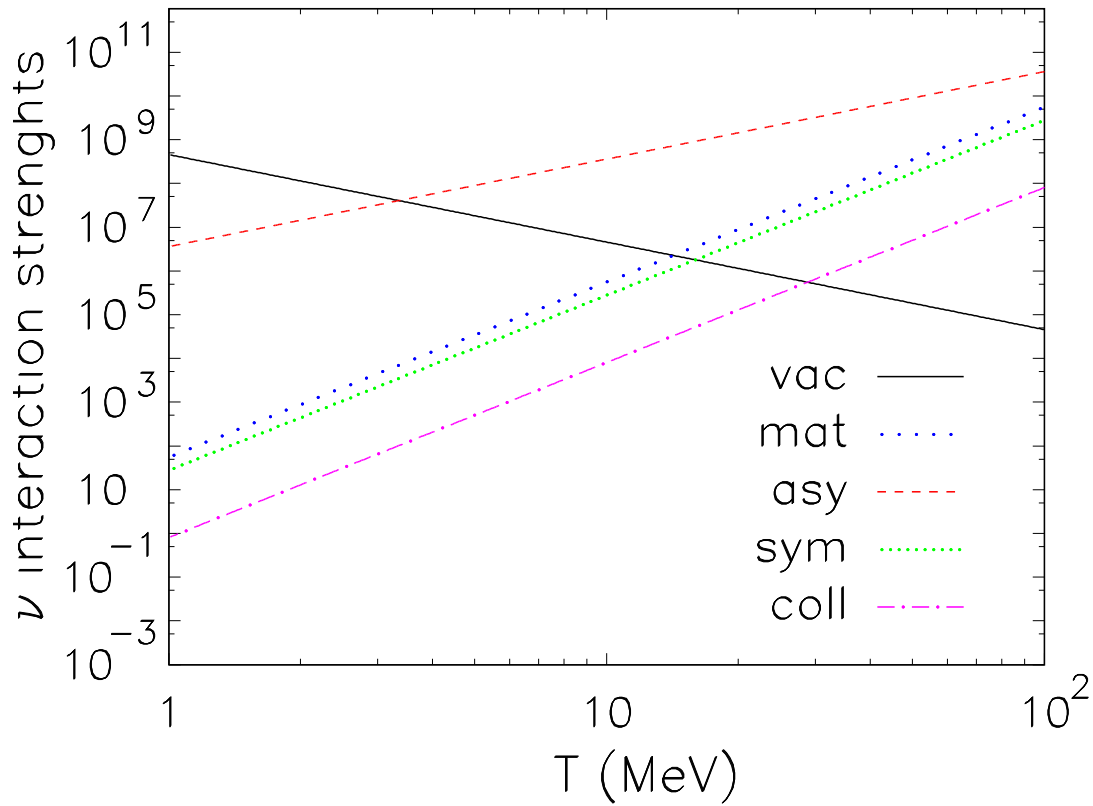} 
 \caption{Neutrino collisional and refractive rates (normalized in terms of the Hubble rate) vs. temperature $T$. In particular, we show  $\Omega_{\rm vac}$ (solid curve), $\Omega_{\rm matt}$ (long-dotted curve),  $\Omega_{\rm asy} \times \Delta_e$ (dashed curve), $\Omega_{\rm sym} \times \Sigma_{ee}$
 (short-dotted curve), 
$\Omega_{\rm coll} \times [(g_s^e)^2 + (g_s^\mu)^2]$  (dash-dotted curve). Here we use $\Delta m^2_{\rm st}$, $\Delta_e= 32 L/3$ with $L=10^{-4}$ and
 $\Sigma_{ee}=2$ (see the text for more details on these quantities).
\label{fig1}}  
\end{figure}  

From the Figure above, one realizes that the system remains collisional down to a few MeV, when the collision over  Hubble rate drops below 1 on the ordinates. The collisional term also dominates over the vacuum oscillation term at  $T\gtrsim 20$~MeV, thus breaking the coherence between different  neutrino flavors and preventing significant oscillations. 
The refractive terms can induce  MSW-like resonances between the actives ($a=e,\mu,\tau$) and sterile state 
when, in the limit of only one mixing angle between the active and the sterile neutrinos,  one of the following conditions is satisfied~\cite{Bell:1998ds}
\begin{eqnarray}
\Omega_{\rm vac} \cos 2 \theta_{a s} -  \Omega_{\rm asy} \Delta_{a} +
\Omega_{\rm sym} \Sigma_{aa} + \Omega_{\rm mat}  &=& 0 \,\ \,\ \textrm{for} \,\ \nu \,\ , \nonumber \\
\Omega_{\rm vac} \cos 2 \theta_{a s} +  \Omega_{\rm asy} \Delta_{a} +
\Omega_{\rm sym} \Sigma_{aa} + \Omega_{\rm mat}  &=& 0  \,\ \,\ \textrm{for} \,\ \overline\nu \,\ ,
\label{eq:res}
\end{eqnarray}
where $a=e,\mu,\tau$ and  the definitions of $\Delta_{\mu}$, $\Delta_{\tau}$ and
 $\Sigma_{\mu\mu}$, $\Sigma_{\tau \tau}$ are respectively of the same form to the ones of 
 $\Delta_{e}$ and  $\Sigma_{ee}$ given before. 
From these equations, we obtain that in absence of lepton asymmetries ($\Delta_a=0$)
the resonance condition cannot be satisfied neither for the $\nu$'s nor for $\overline\nu$'s, given the hypothesis
that the sterile state is heavier than the active ones. 
Instead, when $\Omega_{\rm asy} \Delta_{a}$ is the dominant term, as in the cases we will consider 
in the following, resonance conditions can occur for $\Delta_a>0$ in the $\nu$ sector and for 
 $\Delta_a<0$ in the $\overline\nu$ one. 
In particular, in Fig.~\ref{fig1} the resonance  occurs around $T\simeq 3$~MeV. We will also show that,
as a consequence of the dynamical nature of the asymmetries, $\Delta_a$ can rapidly change sign so
that {\it both} sterile neutrinos and antineutrinos get populated. This phenomenon is thus qualitatively
different with respect to the familiar MSW resonant conversion.  
Resonances can  also take place in the active sector at lower temperatures. However, since active neutrino
distributions are expected not  to  depart too much from their equilibrium values,  their effect is 
sub-leading.

\section{(3+1) results}
 
\begin{figure}[!t]  
\includegraphics[angle=0,width=0.8\columnwidth]{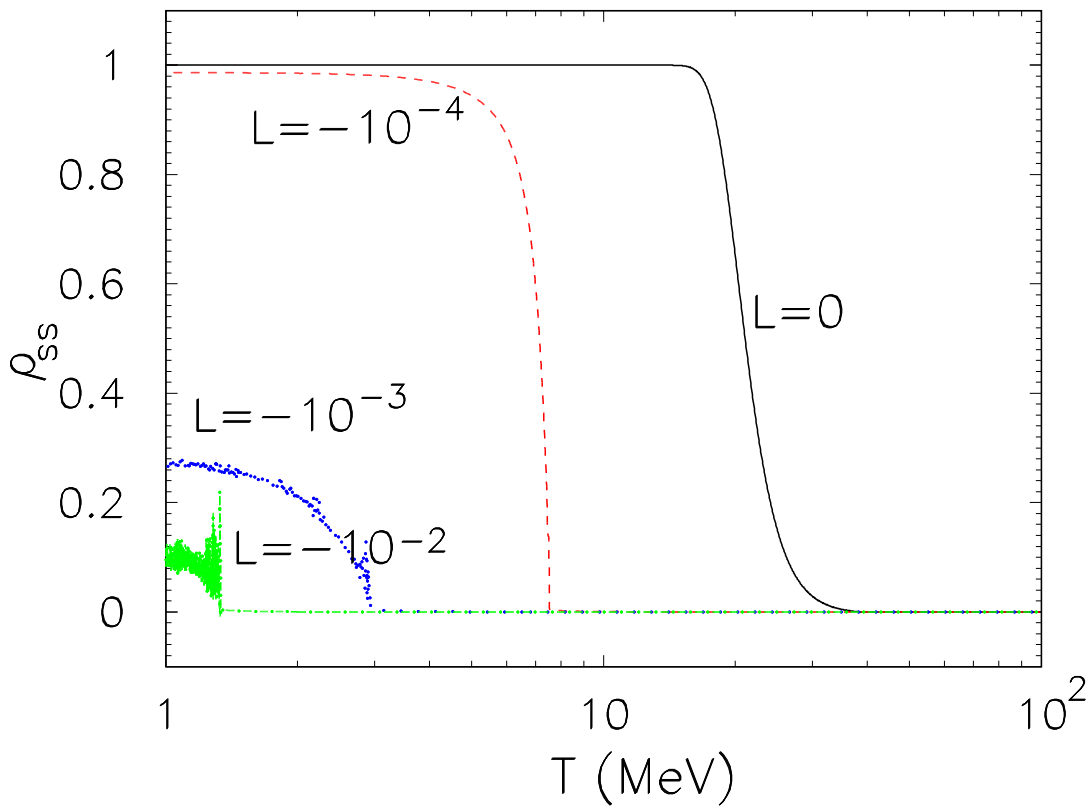} 
 \caption{(3+1) scenario. Evolution of the density matrix element $\rho_{ss}$ in function of the temperature $T$.
 We consider $L=L_e=L_\mu=L_{\tau}$. The solid curve corresponds
 to $L=0$, the dashed curve to $L=-10^{-4}$, the dotted curve to  $L=-10^{-3}$ and the dash-dotted one 
 to $L=-10^{-2}$.
\label{fig2}}  
\end{figure}

In order to calculate the sterile neutrino abundance in the 3+1 scenario,  described in
Sec.~II, we numerically solved the EoMs [Eq.~(\ref{eq:eombarrho_bis})],  using a Runge-Kutta method
for the equations written in the variable $x=m/T$ and evolved  in the range $x \in [10^{-2},1.0]$.  We take $10^5$ steps
in $\log(x)$ in the integration interval. 
We consider  initial neutrino asymmetries $L=L_e=L_\mu=L_{\tau} <0$. We checked that the  results presented in the
following do not change considering  
positive asymmetries. 
In Fig.~\ref{fig2} we show the evolution of the diagonal component of the density matrix $\rho_{ss}$  for
sterile neutrinos  in function of the temperature $T$ for different initial lepton 
asymmetries, namely $L=0$ (solid curve), $L=-10^{-4}$ (dashed curve), 
$L=-10^{-3}$ (dotted curve), and $L=-10^{-2}$ (dash-dotted curve). 
As expected from the previous literature, in absence of lepton asymmetries sterile neutrinos are copiously
produced at $T \lesssim 30$~MeV until they reach $\rho_{ss}=1$. Instead, including a non-zero initial
lepton asymmetry the effect is to suppress the sterile neutrino production as long as $|\Omega_{\rm asy}| \gg
|\Omega_{\rm vac}|$. However, these two functions have opposite dependence on the temperature and
at some time they will cross. Sterile neutrinos are then produced ``resonantly'', albeit with a non-linear, dynamical resonance condition
which is itself influenced by the evolution of the system.  Increasing  the lepton number asymmetry the position of the resonance moves towards
lower temperatures, where  the resonance is less adiabatic. Indeed, 
 the adiabaticity parameter scales as $\sim T$, as shown in~\cite{DiBari:2001jk}. As a consequence, the sterile
production is less efficient  increasing $|L|$, as results from Fig.~\ref{fig2}. 
In particular, asymmetries greater  than $|L |=10^{-3}$ are required in order to achieve a significant 
suppression of the sterile neutrino production. Also, the asymmetric term changes sign and thus the resonance can take place in
both neutrino and antineutrino sectors, which turn out to be populated almost equally.

\begin{figure}[!t]  
\includegraphics[angle=0,width=0.8\columnwidth]{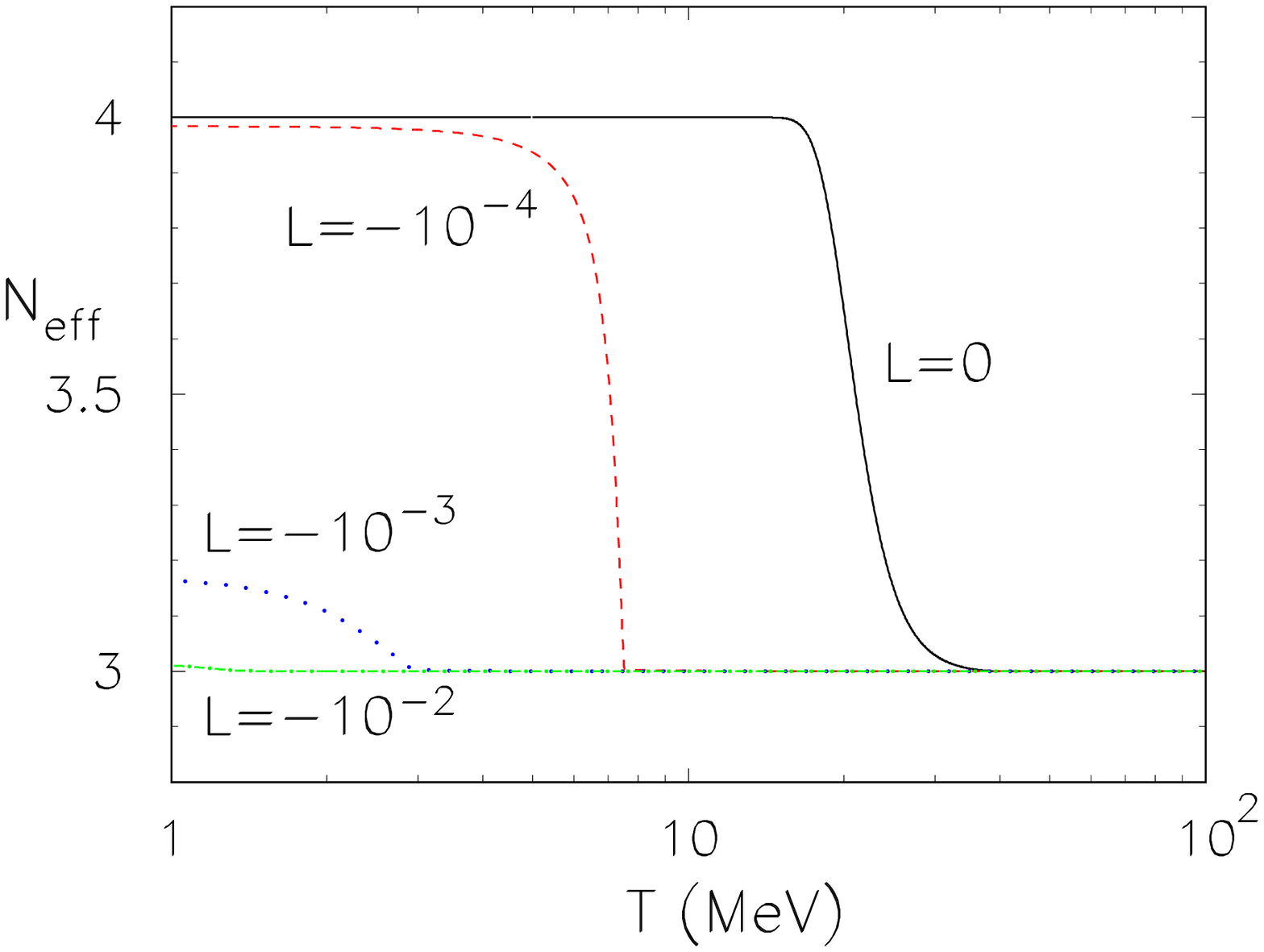} 
 \caption{(3+1) scenario. Evolution of the effective number of degrees of freedom $N_{\rm eff}$ for the cases corresponding to
 Fig.~\ref{fig2} 
\label{temp}}  
\end{figure}  

Our result implies that in order to suppress the sterile neutrino production one needs
 a lepton  asymmetry greater at least by an order of magnitude with respect
to what found in a previous study on the subject~\cite{Chu:2006ua}. 
This discrepancy is due to the fact that in their  work the authors followed the flavor evolution only
for the neutrinos, choosing a negative value of lepton asymmetry kept constant.
In this way, they missed resonant effects that would have occurred in the antineutrino sector. Then, the lepton number can only
suppress  the flavor evolution. Therefore, in their study  $L =-10^{-4}$ was enough to block the sterile neutrino production. 
   
Caution should also be taken when interpreting the results shown for $\rho_{ss}$ into an effective increase
of the neutrino degrees of freedom in the early universe, usually parameterized via $N_{\rm eff}$. In fact, according to the definition reported in Eq.~(\ref{neff}),
the latter variable is sensitive to the trace of the neutrino plus antineutrino density matrix. A late conversion of some active state into a sterile one after
the neutrinos have undergone collisional decoupling is in fact conserving the overall number of neutrinos (albeit some cosmological consequences, such as those
for BBN, may be typically more dramatic, as briefly discussed in Sec.~\ref{bbn}). This is shown in Fig.~\ref{temp}, reporting the evolution of $N_{\rm eff}$ for the cases corresponding to
 Fig.~\ref{fig2}. Note that for no or small asymmetry, for the parameters chosen the active-sterile oscillations take place early enough that the depleted active
 states are rapidly repopulated collisionally. Thus $N_{\rm eff}$ effectively increases to 4.  On the other hand, for $|L|=10^{-3}$ the conversion takes place 
 around the decoupling time, and the repopulation is only partial, with a difference between $N_{\rm eff}$ and $\rho_{ss}$ of about 0.1 units (compare  Fig.~\ref{fig2} with  Fig.~\ref{temp}).  Finally, for $|L|=10^{-2}$ only a negligible fraction of the converted active neutrinos are repopulated, despite the fact that about
 10\% of a ``thermal-equivalent'' sterile state has been produced. Since 
 of large asymmetries the temperature at which production starts depends on when the equality  $\Omega_{\rm vac}=\Omega_{\rm asy} \times \Delta_e$ takes place, there is a quite strong dependence of the signatures from the exact values of
 the active neutrino mass, mixing, and the initial value of $|L|$.

\section{(2+1) results}\label{2plus1}

In our study we consider  initial distributions for active neutrinos  close 
to their equilibrium ones.
Therefore,
the oscillations among the three active species have a sub-leading role for the evolution of the sterile
neutrinos.
At this regard, we calculated the flavor conversions in the same cases as before, considering (2+1) sub-sectors
with the active mixing associated with $(\Delta m^2_{\rm sol}, \theta_{12})$ and 
 $(\Delta m^2_{\rm atm}, \theta_{13})$, respectively. For the cases we compared, we find results very similar to the ones presented
 in the previous section. Therefore,  in order to speed-up the numerical calculations we decide to  continue our explorations 
 of sterile neutrino production in different cases, referring to $(2+1)$ scenarios associated with   $(\Delta m^2_{\rm atm}, \theta_{13})$
 active sector.

\begin{figure}[!t]  
\includegraphics[angle=0,width=0.8\columnwidth]{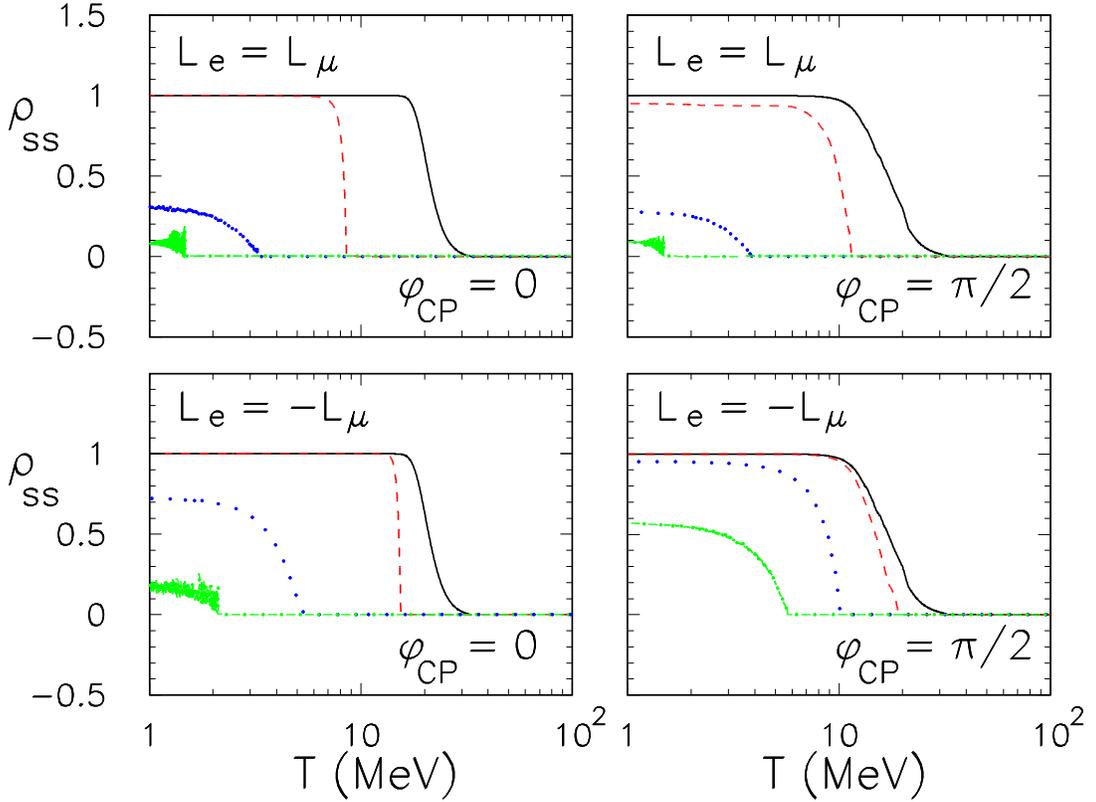} 
 \caption{(2+1) scenario.
Evolution in function of the temperature $T$ of $\rho_{ss}$ for different
initial neutrino asymmetries. Upper panels correspond to $L=L_e=L_{\mu}$, lower
panels correspond to $L=L_e=-L_{\mu}$. The solid curves correspond to
$L=0$, the dashed curves to $L=-10^{-4}$, the dotted to $L=-10^{-3}$ and
the dash-dotted to $L=-10^{-2}$. 
 Left panels show cases with no CP violation
 in the sterile neutrino sector, while  right panels refers to
 $\varphi_{\rm CP}=\pi/2$. 
\label{fig3}}  
\end{figure}  

\subsection{$L_e=L_\mu$,  $\varphi_{CP}=0$ }
\begin{figure}[!t]  
\includegraphics[angle=0,width=0.8\columnwidth]{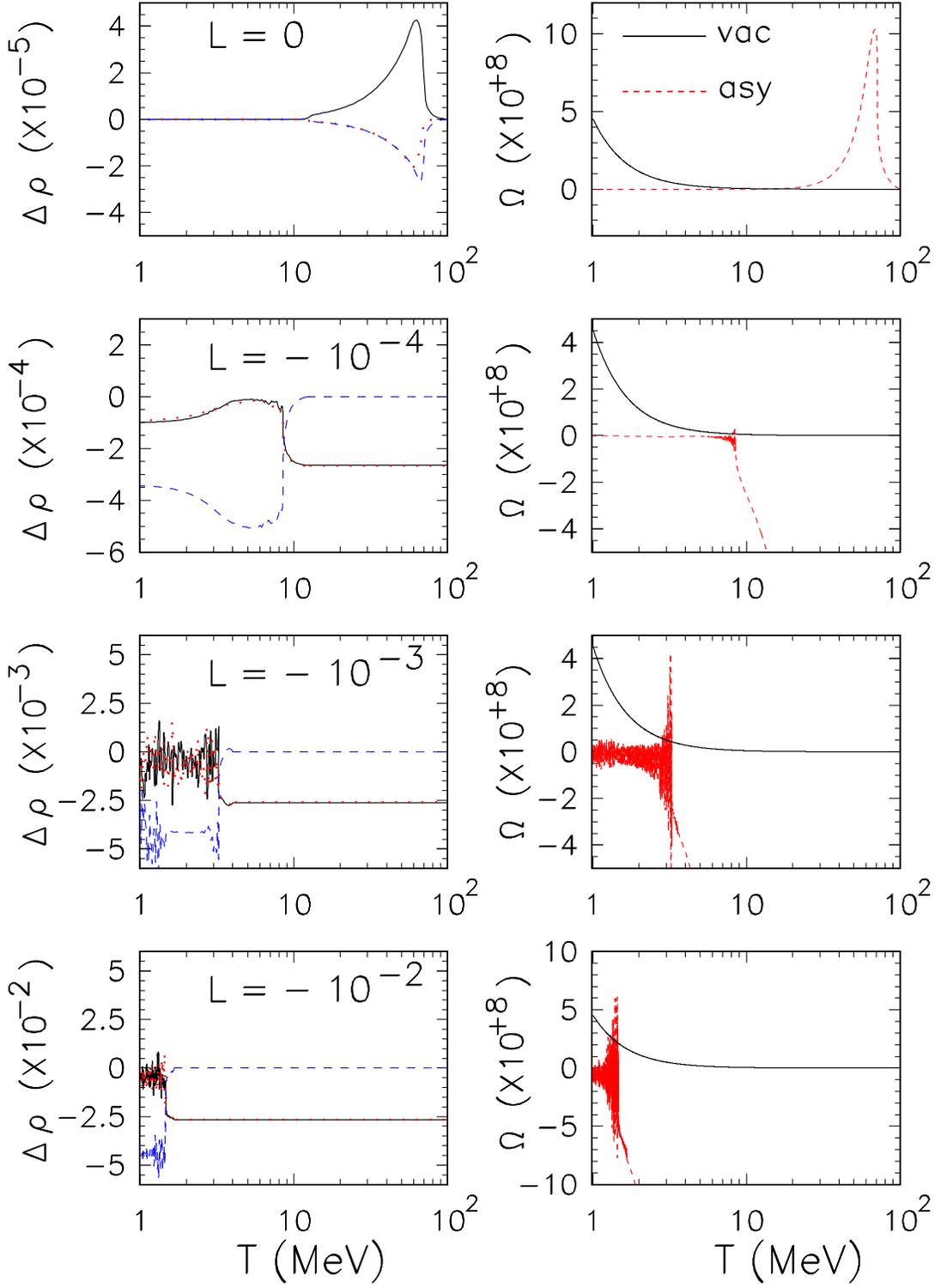} 
 \caption{(2+1) case  with  $L=L_e=L_{\mu}$ and $\varphi_{\rm CP}=0$.
Left panels: Evolution $\Delta \rho_{\alpha}=  \rho_{\alpha\alpha}- \bar\rho_{\alpha\alpha}$
 for the $\nu_e$ (solid curve), $\nu_{\mu}$ (dotted curve) and 
$\nu_s$ (dashed curve) for the different values of initial neutrino asymmetries.
Right panels:  Evolution of $\Omega_{\rm vac}$ (solid curve) vs $\Omega_{\rm asy}\times \Delta_{e}$
(dashed curve).
\label{fig4}}  
\end{figure}

In the following we consider different (2+1) cases with non-zero $\theta_{es}$ and
$\theta_{\mu s}$ given by  Eq.~(\ref{eq:stermix}). 
In the left-upper panel of Fig.~\ref{fig3} we  represent the case with $L=L_e=L_\mu$.
The solid curve corresponds
 to $L=0$, the dashed curve to $L=-10^{-4}$, the dotted curve to  $L=-10^{-3}$ and the dash-dotted one 
 to $L=-10^{-2}$.
 This case is manifestly  close  to the (3+1) scenario shown in Fig.~\ref{fig2}. 
 In order to clarify the dynamics of the sterile neutrino production,  in the left panels of Fig.~\ref{fig4} 
we plot in function of the temperature, the  evolution of the neutrino asymmetries 
 $\Delta \rho_{\alpha} =  \rho_{\alpha\alpha}- \bar\rho_{\alpha\alpha}$ for the $\nu_e$ (solid curve), $\nu_{\mu}$ (dotted curve) and 
$\nu_s$ (dashed curve).
Since $\Delta \rho_{\alpha}$ typically presents very fast oscillations, for the sake of the clarity 
we plot its value averaged over ten steps in $T$.   
 In the right panels
we show  the evolution of the  vacuum term $\Omega_{\rm vac}$ (solid curve) and of the $\Omega_{\rm asy} \times
\Delta_{e}$ term (dashed curve) for the same cases of the left panels. The crossing between these two curves at
non-zero $L$ determines
the position of a $e$-$s$ resonance.

Starting with the case $L=0$, we see that $L_e = -2 L_\mu=- 2 L_s \simeq  \textrm{few} \times 10^{-5}$ can be dynamically  generated
at the onset of the flavor conversions (at $T \lesssim 80$~MeV). Since the active asymmetries are opposite, they tend to decrease reaching flavor 
equilibrium ($L=0$) at $T \simeq 10$~MeV. At $T \lesssim 30$~MeV, when collisional rates slow down enough
(see Fig.~\ref{fig1}),  sterile neutrinos are produced without any 
hindrance (see Fig.~\ref{fig3}).  

We pass now to the cases with non-zero initial neutrino asymmetries. In this situation, 
since $\theta_{es}$ and $\theta_{\mu s}$ are non-vanishing,  both the active states 
can have resonances with the sterile one. Moreover, since for our choice
$\theta_{es}\simeq \theta_{\mu s}$, the evolution of $\Delta \rho_e$  and $\Delta \rho_\mu$ is very similar. 

In the case with initial $L=-10^{-4}$  the production of $\nu_s$ starts
at $T\simeq 10$~MeV (Fig.~\ref{fig3}) when an active-sterile resonance occurs.
Also in the other two cases with $L=-10^{-3}$ and $L=-10^{-2}$ the position of the resonance coincides
with the onset in the rise of $\rho_{ss}$ in Fig.~\ref{fig3}.
However, as commented before, the lower the resonance temperature, the less adiabatic the resonance. 
Therefore, 
 the sterile neutrino production   is further  inhibited.

\subsection{$L_e=L_\mu$,  $\varphi_{CP}=\pi/2$}
Fits for  laboratory anomalies have been proposed which include CP violation effects in
the sterile sector (see e.g.~\cite{Karagiorgi:2006jf}). Perhaps more importantly, whenever
three or more neutrinos mix, CP-violating ``Dirac phases'' entering oscillations are naturally present
in the theory. For both reasons, we find worthwhile to investigate the impact of CP-violation in our framework.
 
\begin{figure}[!t]  
\includegraphics[angle=0,width=0.8\columnwidth]{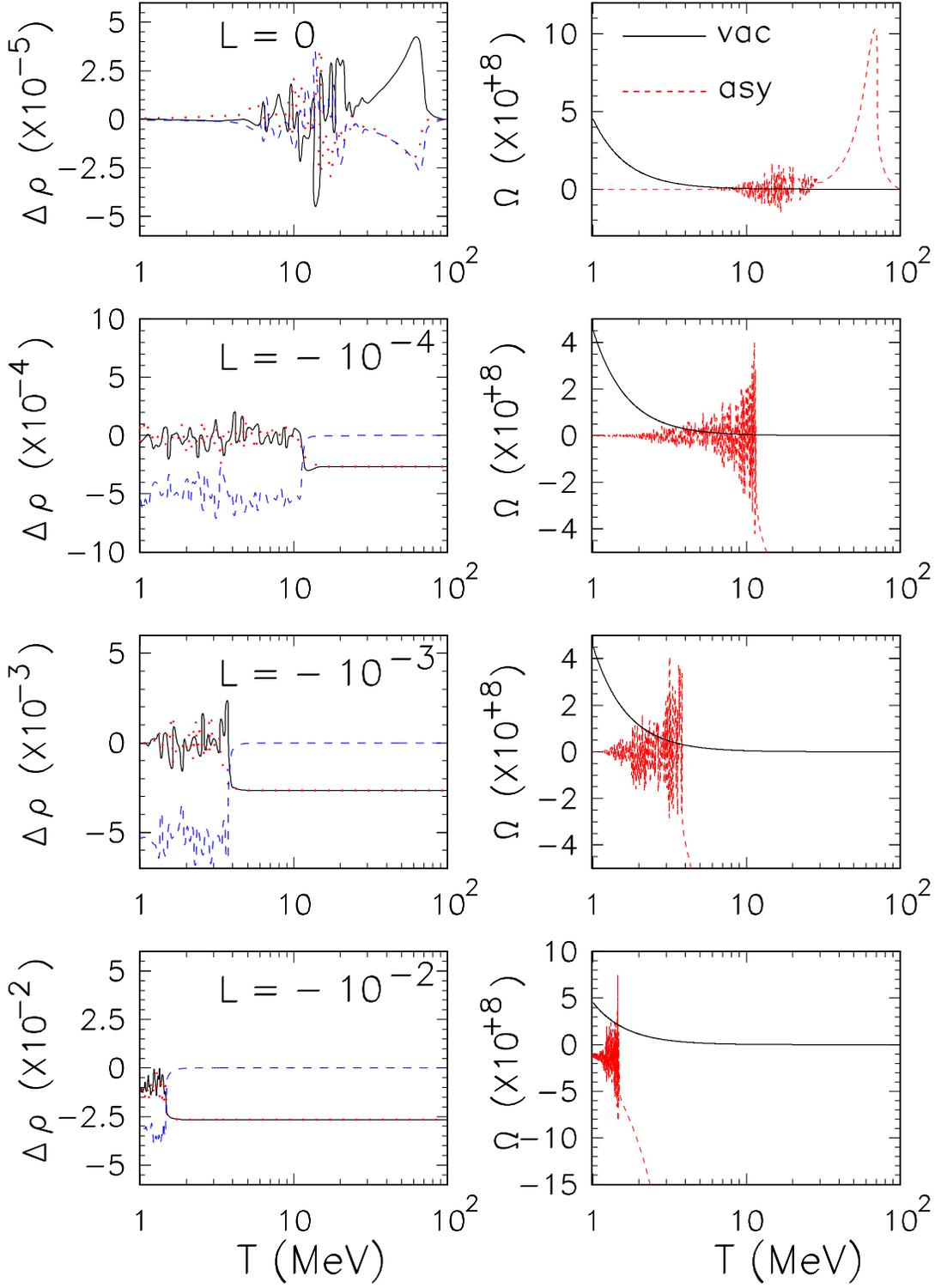} 
 \caption{(2+1) case  with  $L=L_e=L_{\mu}$ and $\varphi_{\rm CP}=\pi/2$.
Left panels: Evolution $\Delta \rho_{\alpha}=  \rho_{\alpha\alpha}- \bar\rho_{\alpha\alpha}$
 for the $\nu_e$ (solid curve), $\nu_{\mu}$ (dotted curve) and 
$\nu_s$ (dashed curve) for the different values of initial neutrino asymmetries.
Right panels:  Evolution of $\Omega_{\rm vac}$ (solid curve) vs $\Omega_{\rm asy}\times \Delta_{e}$
(dashed curve).
\label{fig5}}  
\end{figure}

For this purpose, we include 
an extra phase in the sterile-active  mixing matrix [Eq.~(\ref{eq:matrix})], formally in  the same way the Dirac
phase is introduced in $3\times 3$ active neutrino mixing formalism.
The inclusion of CP violating effects in the sterile sector could be potentially interesting,
since it would generate an asymmetry among sterile neutrinos and antineutrinos. This  could be transferred by oscillations 
into the 
active sector, having a feedback on the further growth of the sterile neutrino abundance. 
For definiteness we consider  $\varphi_{CP}=\pi/2$.  Note also that in the full (3+1) scenario---not to speak 
of the (3+2) scenarios with 2 sterile states---the number of CP-violating phases grows. Therefore, the present investigation
is expected  to be conservative in some respect.
We first refer to the case with initial equal neutrino asymmetries among active species:  
$L=L_e=L_{\mu}$. The evolution of $\rho_{ss}$ is shown in the right-upper panel of Fig.~\ref{fig3}, whose  comparison with the CP conserving case
shows that the suppression of the sterile neutrino abundance 
due to $\varphi_{CP}$   is sub-leading.   
Indeed, from Fig.~\ref{fig5} one sees that the growth of the dynamical neutrino asymmetries $\Delta \rho_{\alpha}$
for different intial $L$, even if it is more irregular than in the case 
with $\varphi_{\rm CP}=0$ (Fig.~\ref{fig4}), it is qualitatively similar. This  implies that this
effect does not significantly alter the flavor evolution.  On the other hand, it is interesting to note that even in absence
of an initial neutrino asymmetry the CP-violating mixing can create a ``dynamical'' asymmetry at relatively late times (down to
decoupling temperatures) which is of the order to $10^{-5}$  for the parameters used. 

\subsection{$L_e=-L_\mu$,  $\varphi_{CP}=0$}

\begin{figure}[!t]  
\includegraphics[angle=0,width=0.8\columnwidth]{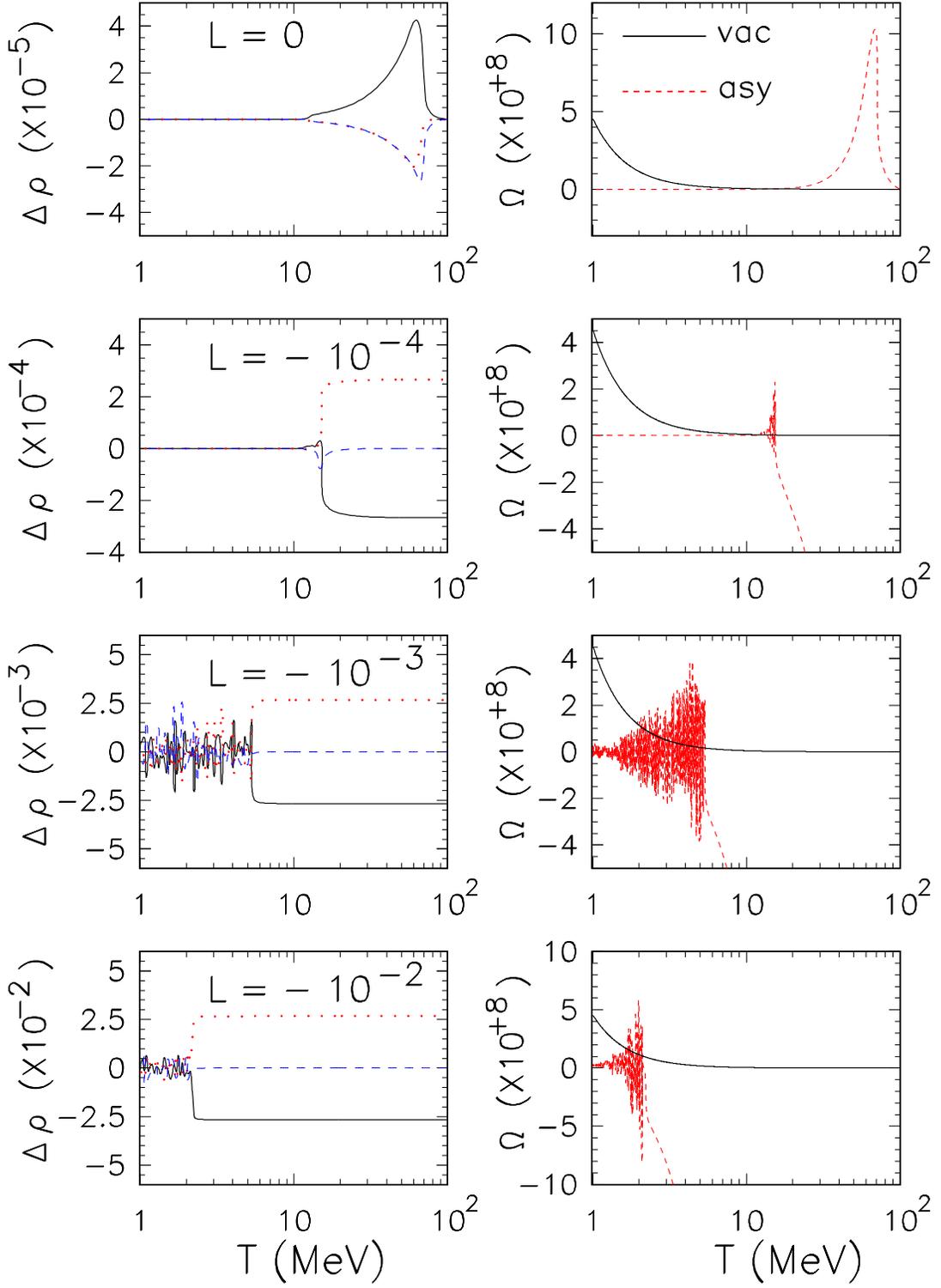} 
 \caption{(2+1) case  with  $L=L_e=-L_{\mu}$ and $\varphi_{\rm CP}=0$.
Left panels: Evolution $\Delta \rho_{\alpha}=  \rho_{\alpha\alpha}- \bar\rho_{\alpha\alpha}$
 for the $\nu_e$ (solid curve), $\nu_{\mu}$ (dotted curve) and 
$\nu_s$ (dashed curve) for the different values of initial neutrino asymmetries.
Right panels:  Evolution of $\Omega_{\rm vac}$ (solid curve) vs $\Omega_{\rm asy}\times \Delta_{e}$
(dashed curve).
\label{fig6}}  
\end{figure}

We pass now to consider the case in which the initial neutrino asymmetries in the active sector are opposite
for $\nu_e$ and $\nu_{\mu}$, i.e. $L=L_e=-L_{\mu}$. In the absence of CP violation, this case is represented
in the bottom-left panel of Fig.~\ref{fig3}. 
For a non-vanishing initial $L$  the sterile neutrino production is enhanced
with respect to the previous case with equal  asymmetries among the different flavors. Indeed, to achieve
a significant suppression of the sterile species one needs an initial $|L|\sim 10^{-2}$, i.e. roughly one order
of magnitude larger than in the previous case.  
This behavior can be clarified looking at the evolution of the dynamical  asymmetries $\Delta \rho_{\alpha}$
shown for the different initial $L$ in Fig.~\ref{fig6}. 
We remark that since $\Delta \rho_{e}$ and $\Delta \rho_{\mu}$ have opposite sign,  resonances can occur
 simultaneously  in the neutrino and antineutrino channels.
When
these happen, they tend to produce  flavor equilibrium between $\nu_e$ and $\nu_{\mu}$. This leads to
 a vanishing final lepton number. When the neutrino asymmetry is destroyed, the sterile neutrinos can be produced
without any hindrance. This explains the enhancement  in the final $\rho_{ss}$ found in this case.

\subsection{$L_e=-L_\mu$,  $\varphi_{CP}=\pi/2$}

\begin{figure}[!t]  
\includegraphics[angle=0,width=0.8\columnwidth]{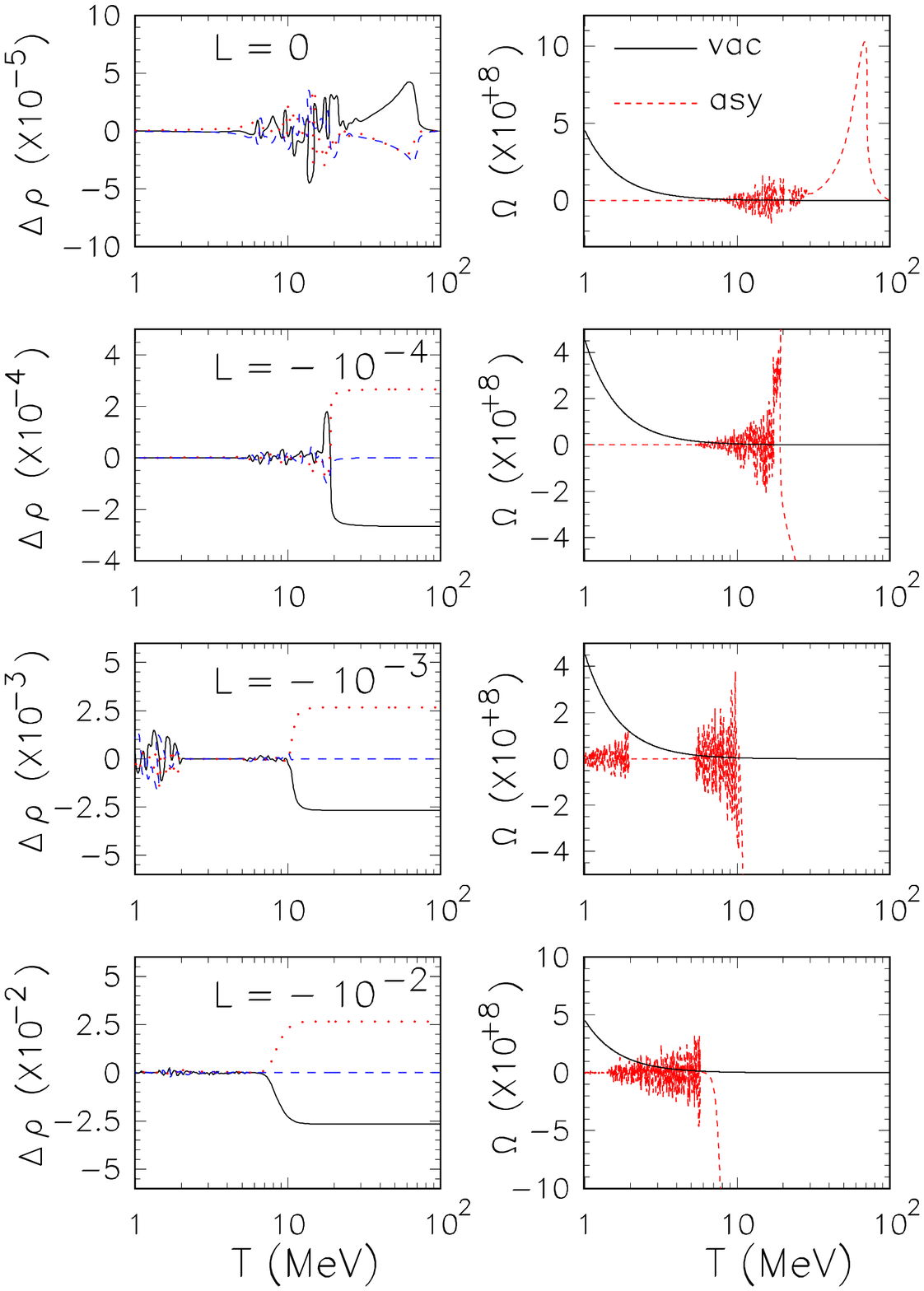} 
 \caption{(2+1) case  with  $L=L_e=-L_{\mu}$ and $\varphi_{\rm CP}=\pi/2$.
Left panel: Evolution $\Delta \rho_{\alpha}=  \rho_{\alpha\alpha}- \bar\rho_{\alpha\alpha}$
 for the $\nu_e$ (solid curve), $\nu_{\mu}$ (dotted curve) and 
$\nu_s$ (dashed curve) for the different values of initial neutrino asymmetries.
Right panel:  Evolution of $\Omega_{\rm vac}$ (solid curve) vs $\Omega_{\rm asy}\times \Delta_{e}$
(dashed curve).
\label{fig7}}  
\end{figure}

We now consider the case with opposite initial neutrino asymmetries and  $\varphi_{\rm CP}=\pi/2$. The evolution of $\rho_{ss}$ in this case is shown 
in the bottom-right panel of Fig.~\ref{fig3}. 
The production of sterile neutrinos is significantly enhanced with respect to the previous cases. 
In particular, also for an initial $L=-10^{-2}$, the final abundance of sterile neutrinos is relevant. 
From Fig.~\ref{fig7} one sees that the flavor equilibrium  between the electron and muon species
occurs at higher $T$ than in the CP conserving case (Fig.~\ref{fig6}).
Indeed, CP violating effects tend to create an asymmetry in the sterile sector. This  would push the active system
earlier to equilibrium in order to conserve the total null neutrino asymmetry. Since $L$ is equilibrated at higher
temperature with respect to the CP conserving case, sterile neutrinos are produced more efficiently.

\subsection{$\theta_{\mu s}=0$}

\begin{figure}[!t]  
\includegraphics[angle=0,width=0.8\columnwidth]{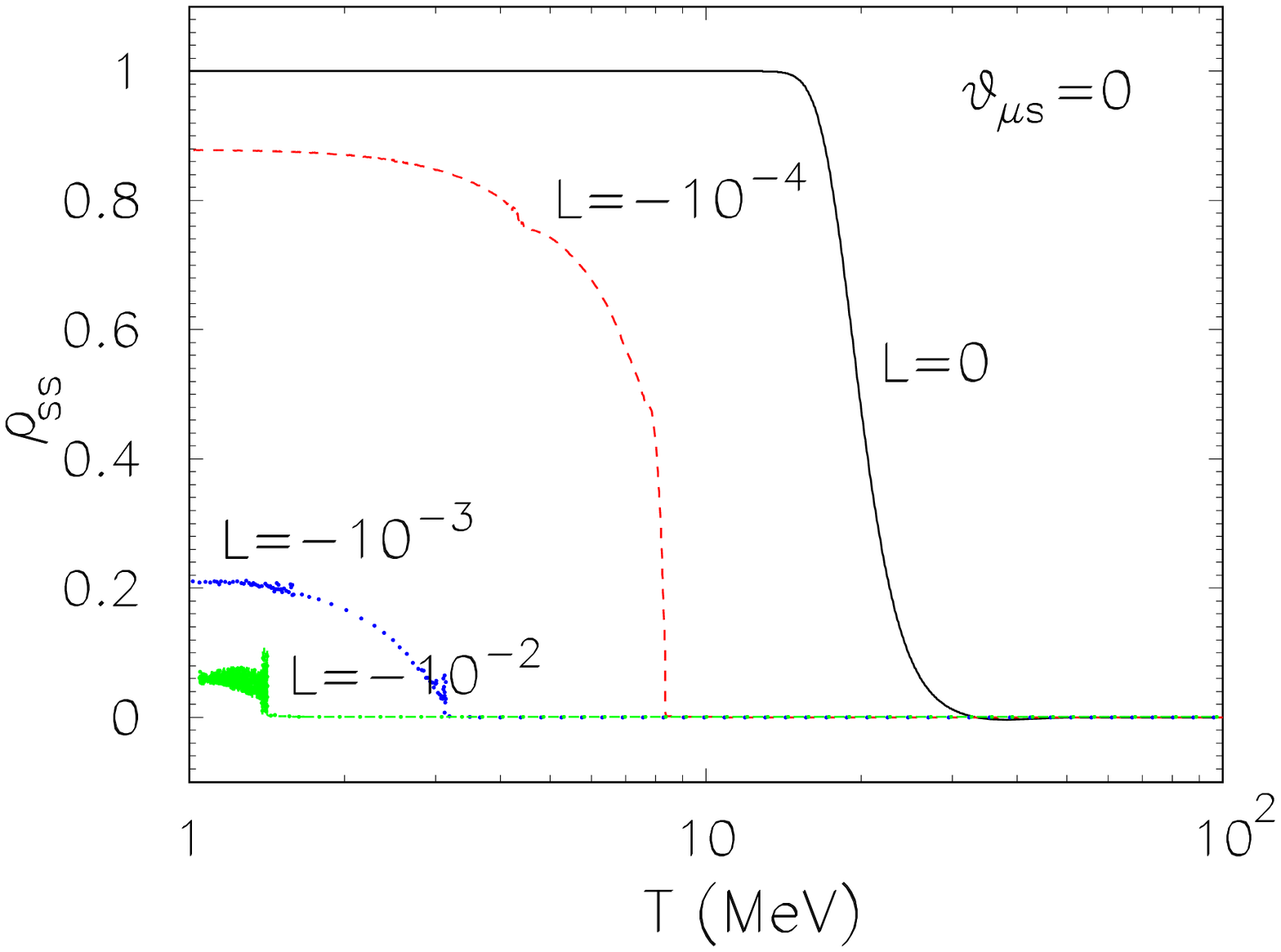} 
 \caption{(2+1) case with $\theta_{\mu s}=0$. Evolution of the density matrix element $\rho_{ss}$ in function of the temperature $T$.
 We consider $L=L_e=L_{\mu}$. The solid curve corresponds
 to $L=0$, the dashed curve to $L=-10^{-4}$, the dotted curve to  $L=-10^{-3}$ and the dash-dotted one 
 to $L=-10^{-2}$.
\label{fig8}}  
\end{figure}  

In the recent literature,  models in which sterile neutrinos mix only with the (mostly) electron
 ones have been discussed as well~\cite{Kopp:2011qd}.
Hence we also consider a $(2+1)$ case in which the mixing angle $\theta_{\mu s}=0$, while
$\theta_{e s}$ is given by Eq.~(\ref{eq:stermix}). 
The evolution of the sterile neutrino abundance $\rho_{ss}$ in function of $T$ is shown in Fig.~\ref{fig8}.
 For the sake of the brevity, we only consider  equal initial neutrino asymmetries $L=L_e=L_\mu <0$. We represent the cases  $L=0$ (solid curve), $L=-10^{-4}$ (dashed curve), 
$L=-10^{-3}$ (dotted curve), and $L=-10^{-2}$ (dash-dotted curve). 
From the comparison with the analogous (2+1) case with two active-sterile mixing angles (Fig.~\ref{fig3}),
 we see that the evolution of $\rho_{ss}$ at
different $L$ is qualitatively similar.  However, for non-zero  asymmetries  the production
of sterile neutrinos is slightly suppressed with respect to the two mixing scenario.

\begin{figure}[!t]  
\includegraphics[angle=0,width=0.8\columnwidth]{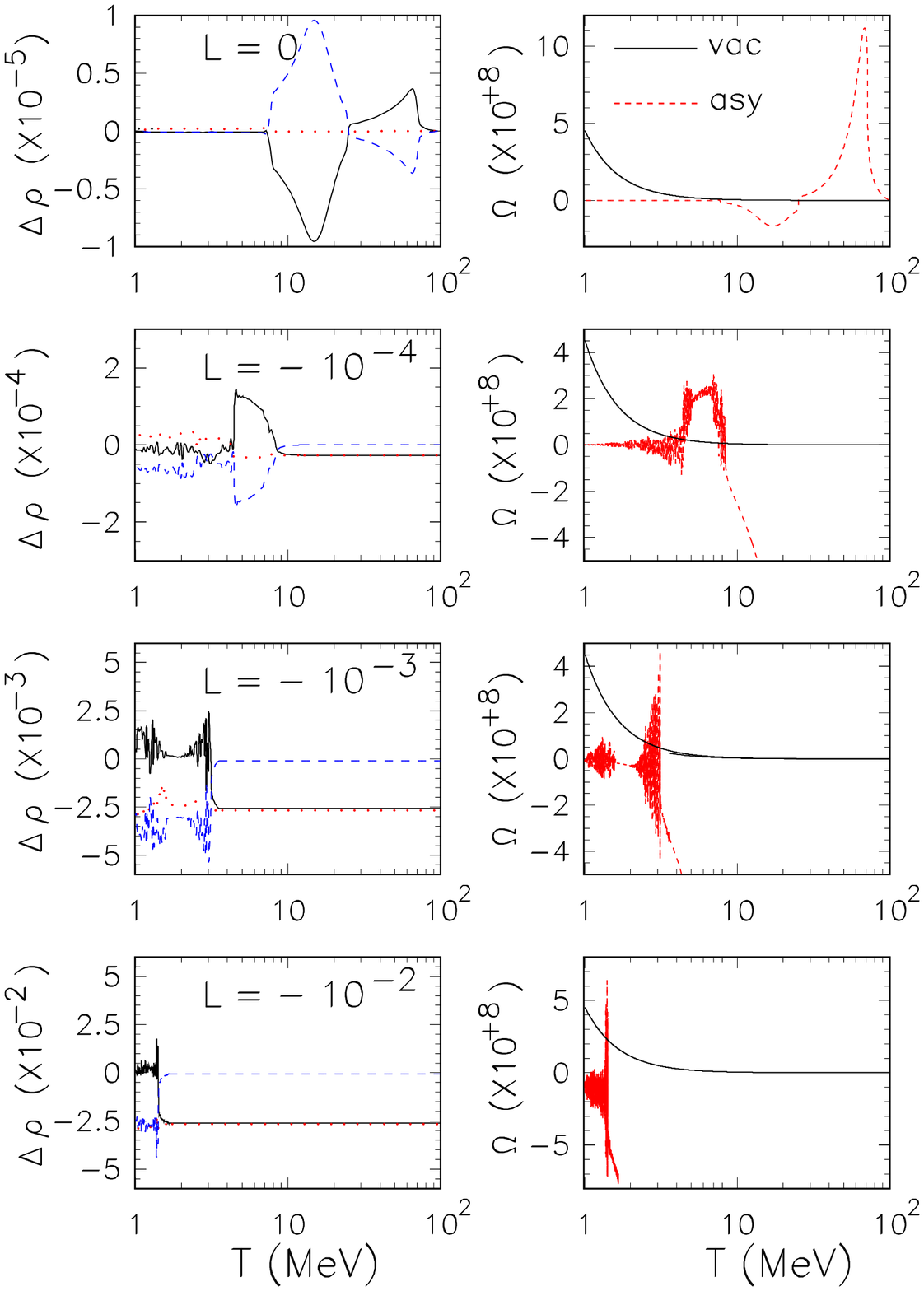} 
 \caption{(2+1) case  with $\theta_{\mu s}=0$ and $L=L_e=L_{\mu}$.
Left panels: Evolution $\Delta \rho_{\alpha}=  \rho_{\alpha\alpha}- \bar\rho_{\alpha\alpha}$
 for the $\nu_e$ (solid curve), $\nu_{\mu}$ (dotted curve) and 
$\nu_s$ (dashed curve) for the different values of initial neutrino asymmetries.
Right panels:  Evolution of $\Omega_{\rm vac}$ (solid curve) vs $\Omega_{\rm asy}\times \Delta_{e}$
(dashed curve).
\label{fig9}}  
\end{figure}  

As in the previous cases, in  Fig.~\ref{fig9} 
we plot  the evolution of the asymmetries  $\Delta \rho_{\alpha}$ for
 the different values of initial $L$.
In the case $L=0$,   a value 
$L_e = -L_s \simeq  \textrm{few} \times 10^{-5}$ can be dynamically  generated.
Since $\nu_\mu$'s  are not mixed with the sterile states,  their asymmetry remains identically
zero.  
The positive  $L_e$
 can generate  a $\nu_e$-$\nu_s$ resonance in the neutrino sector at 
$T\simeq 30$~MeV. Then, $L_e$ becomes negative reaching a value $L_e \simeq -10^{-4}$.
After that, the electron and the sterile neutrinos go towards flavor equilibrium (with $L=0$), reaching 
it at $T\simeq 10$~MeV.
In the case with initial $L=-10^{-4}$ the production of $\nu_s$ starts
at $T\simeq 10$~MeV when a $\nu_e$-$\nu_s$ resonance occurs. We note that since $\Omega_{\rm asy}\times\Delta_{e}$
is a rapidly oscillating function taking both positive and negative values, resonances affect both neutrino
and antineutrino channels. 
 Then, $L_e$ reaches  a value $\sim 2\times 10^{-3}$, while $L_\mu \simeq 0$. A second series of  resonances occurs at $T\simeq 4$~MeV, 
leading $L_e$ to zero and $L_s\simeq -10^{-4}$.
In the last two cases, only one resonance occurs at $T\simeq 3$~MeV for $L=-10^{-3}$,
and   $T\simeq 0.5$~MeV for  $L=-10^{-2}$. Once more the sterile neutrino production is triggered
by these resonances.

\section{Semi-analytical estimate of the effects on BBN}\label{bbn}
In order to compute in detail the effects of adding a fourth, sterile neutrino onto BBN, full  momentum-dependent  
calculations are necessary. This is essentially due to the fact that the $\nu_e$ and $\overline\nu_e$ distributions
enter the weak rates regulating the neutron-proton equilibrium and, eventually, the amount of surviving neutrons
which will mostly end up bound in $^4$He nuclei (plus traces of some other light elements), see~\cite{Steigman:2007xt,Iocco:2008va}
for reviews. We can however provide a crude estimate based on a simple physical argument, which has already been used
before in this context (see for example~\cite{Dolgov:2003sg}).  Modifying the neutrino sector alters both the Hubble expansion
rate and the overall magnitude of the isospin-changing weak rates $\Gamma_{\rm iso}$, the latter through a change of
the neutrino and antineutrino number density parameterized here by $\rho_{ee}$~\footnote{For the present considerations we assume $\rho_{ee}=\overline\rho_{ee}$ and
neglect the further effect of unbalancing $n\to p$ vs. $p\to n$ rates due to asymmetries $L$, as well as the modified contribution to the Hubble rate due
to the asymmetries. For all cases considered here they produce only sub-leading changes compared to those illustrated in this section.}. Hence the freeze-out temperature $T_F$, as defined by the condition
\begin{equation}
\Gamma_{\rm iso}(\rho_{ee},T_F)=H(N_{\rm eff},T_F)\,,
\end{equation} 
is altered with respect to its standard value $T_F\simeq 0.8\,$MeV due to a higher-than-standard $N_{\rm eff}$ and a lower-than-standard $\rho_{ee}$. Both effects go in the direction of increasing $T_F$ and, as a consequence,  anticipate the freeze-out of $n/p$. This ratio is lower then unity due to the fact that neutrons are heavier than protons by $Q=1.293\,$MeV, a non-negligible
quantity compared to the energies involved when $T$ drops to the MeV scale.
Since the $^4$He mass abundance $Y_p$ is proportional to the $n/p$ ratio at freeze-out
\begin{equation}
Y_p \propto \left(\frac{n}{p}\right)_{T_F} \propto e^{-Q/T_F}\,,
\end{equation}
we obtain the estimate 
\begin{equation}
\frac{\delta Y_p}{Y_p}=\frac{Q}{T_F}\frac{\delta T_F}{T_F}\simeq 1.6 \frac{\delta T_F}{T_F}\,,
\end{equation}
which confirms that we expect an increase in the produced yield. A simple estimate~\cite{Dolgov:2003sg} for the Hubble parameter $H\propto \sqrt{22/7+ N_{\rm eff}}\,T^2$ and the
weak rates $\Gamma_{\rm iso}\propto (1+\rho_{ee})\,T^5$ predicts
\begin{equation}
T_F\propto\left(\frac{\sqrt{22/7+ N_{\rm eff}}}{1+\rho_{ee}}\right)^{1/3}\,,
\end{equation}
which immediately illustrates why $Y_p$ is comparatively much more sensitive to alteration of the weak rates than to the expansion rate via $N_{\rm eff}$. A perturbative expansion
around the fiducial values $N_{\rm eff}=3$ and $\rho_{ee}=1$ yields 
\begin{equation}
\frac{\delta T_F}{T_F}=0.027\,\delta N_{\rm eff}-0.17\,\delta \rho_{ee} \Longrightarrow \frac{\delta Y_p}{Y_p} = 0.044\, \delta N_{\rm eff}-0.27\, \delta\rho_{ee}\,.
\end{equation}

For specific cases suggested by our previous analysis, e.g. for the (3+1) results of Sec.~IV, one finds for example that for $L=0$, $\delta \rho_{ee}\simeq 0$ while $\delta N_{\rm eff}\simeq 1$, hence one deduces a variation in the Helium content of $4.4\%$, which is a large number and barely allowed (see e.g.~\cite{Mangano:2011ar}).  For the largest asymmetries
we considered, $|L|=10^{-2}$, the variation in $N_{\rm eff}$ is negligible while $\delta\rho_{ee}\simeq -0.05$, implying again a few percent effects on $Y_p$. For intermediate values like $|L|=10^{-3}$, one expects again effects above the 1\% level, this time with both terms contributing.
Note that such effects are larger than theoretical uncertainties and comparable to observational ones, hence they do imply that sterile neutrinos cannot be ``easily masked'' to BBN: they do have an impact that must be accounted for in any realistic analysis combining cosmological observables. The above estimates should be considered only as illustrative,
lacking a proper account of momentum-dependent effects in weak rates. We plan to provide a more reliable estimate
of their impact in a future work. 

\section{Conclusions}

Light sub-eV sterile neutrinos,  suggested to solve different anomalies in short-baseline, reactor and
solar experiments~\cite{Akhmedov:2010vy,Kopp:2011qd,Giunti:2011gz,Giunti:2011hn}, could play an interesting cosmological role providing the amount of extra-radiation indicated
by different cosmological observations~\cite{Hamann:2010bk}. However, for the scenarios proposed to fit the different
laboratory data,  sterile neutrinos would be produced too copiously in the early universe by the mixing with the active 
species. This would create a tension between the laboratory hints and the cosmological observations~\cite{Hamann:2011ge}. 
A possibility to reconcile sterile neutrinos with cosmology is the introduction of a primordial neutrino 
asymmetry~\cite{Foot:1995bm}, that
is expected to suppress the sterile-active mixing when its strength dominates over the other interaction terms. 
In this context, we calculated the sterile neutrino abundance in the early universe in (3+1) and (2+1) schemes
 solving the neutrino kinetic equations for different initial asymmetries. 
Considering approximately equilibrium distributions for the active neutrino species, 
  the flavor dynamics of  active neutrinos plays a sub-leading role in determining the final abundance of the sterile
  species. Therefore, (2+1) schemes are a good proxy for the complete (3+1) situation. 
 Starting with initial neutrino asymmetries equal for the two active species, $|L|\simeq 10^{-3}$
would be required to have a significant suppression of the neutrino abundance.
 Otherwise, sterile neutrinos would be produced by resonances between the vacuum term and the evolving (oscillating)  active $\nu-\overline\nu$
 asymmetry potential. Opposite initial neutrino asymmetries (hence a globally vanishing lepton number) would cuase an enhancement in the sterile neutrino production
 compared to the above case,  implying  $|L|\gtrsim 10^{-2}$ to substantially inhibit their creation. 
Moreover, in this last case the presence of CP violating effects would further increase the sterile neutrino abundance, requiring an even larger initial asymmetry to prevent 
their growth. Both the assumptions of non-dynamical asymmetries and, to a minor extent, of mixing with a single active neutrino
tend to underestimate the value of $|L|$ needed for inhibiting the sterile $\nu$ production.

Coming to phenomenological consequences, this suggests that some proposed ways to reconcile ``hints for a large 
$N_{\rm eff}$'' from CMB with
more stringent requirements from BBN, via the introduction of large chemical potentials (see e.g.~\cite{Hamann:2011ge}) are dynamically
hard---if not impossible---to achieve. Whenever CMB feels a large $N_{\rm eff}$ due to  sterile neutrinos of the kind suggested by laboratory anomalies,
BBN should feel the same. However, the opposite situation is not necessarily true. 
Even more interestingly, we found that whenever a significant suppression of the sterile neutrino production takes place thanks to initial asymmetries, the  
active neutrinos have partially or mostly decoupled. This implies that the (small) fraction of them which oscillates into sterile states is not repopulated.
Hence, one expects different possible regimes: For too small asymmetries, $|L|\ll10^{-3}$, the sterile neutrinos are fully populated and their
``parent'' active neutrino spectra are repopulated in the thermal plasma. This implies $N_{\rm eff}\simeq 4$ (for (3+1) scenarios considered here) and
a tension with cosmological mass bounds, which  counteracts the modest fit improvements due to a larger
$N_{\rm eff}$. Increasing the asymmetry ($|L|\gtrsim 10^{-3}$)
the effect on $N_{\rm eff}$ becomes less and less prominent, and completely negligible when $|L|\gtrsim 10^{-2}$. However, the lack of repopulation of electron
neutrinos would in general produce distorted distributions, which can anticipate the $n/p$ freeze-out and hence increase the $^4$He yield, to which BBN is much more sensitive
than CMB. Finally, for a too large $|L|$, no production/depletion takes place, but  these asymmetries in the {\it active flavors} would then become an interesting
cosmological observable to be associated with {\it sterile} neutrinos. 

To go beyond semi-analytical estimates, especially to detail the intermediate regime, one has to relax the average momentum approximation  used in this exploratory study. Due to the momentum-dependence of the resonant conversions between active and sterile neutrinos, a detailed treatment solving the full momentum-dependent equations is necessary to derive quantitative phenomenological predictions. We plan to perform this exploration in a forthcoming article.  
Also note that our study suggests that the dynamics of sterile neutrinos in the early universe is quite dependent on the details of the 
scenario considered. Pinning down the parameters favored by interpretations of the laboratory data in terms of sterile states is crucial
in order to treat in the most accurate way only scenarios which are phenomenologically attractive. Large scans of parameter space obtained
in too crude approximations might miss essential aspects of the problem, which is intrinsically non-linear.

\section*{Acknowledgements} 

We thank Gianpiero Mangano and Ofelia Pisanti for interesting discussions during the
development of this project and Marco Cirelli and Irene Tamborra for comments on the 
manuscript. 
The work of A.M. and N.S.  was supported by the German Science Foundation (DFG)
within the Collaborative Research Center 676 ``Particles, Strings and the
Early Universe''.
G.M. acknowledges support by
the Istituto Nazionale di Fisica Nucleare I.S. FA51 and the PRIN 2010  ``Fisica
Astroparticellare: Neutrini ed Universo Primordiale'' of the Italian Ministero
dell'Istruzione, Universit{\`a}  e Ricerca.

\section*{References} 


\end{document}